\documentclass[12pt,preprint]{aastex}
\renewcommand{\farcs}     {\mbox{\ensuremath{.\!\!^{\prime\prime}}}}

\renewcommand{\ion}[2]{#1$\;${\scshape{#2}}}

\newcommand{\hii}{\mbox{\ion{H}{ii}}}
\newcommand{\nev}{\mbox{[\ion{Ne}{v}]}}
\newcommand{\ha}{\mbox{H$\alpha$}}

\shortauthors{Desroches \& Ho}
\shorttitle{Candidate AGNs in Late-Type Galaxies}

\begin{document}


\slugcomment{\today; accepted to the {\it Astrophysical Journal}}

\title{Candidate Active Nuclei in Late-type Spiral Galaxies}
\author{Louis-Benoit Desroches\altaffilmark{1} and Luis C. Ho\altaffilmark{2}}

\altaffiltext{1}{Department of Astronomy, University of California, Berkeley,
CA 94720-3411; louis@astro.berkeley.edu}
\altaffiltext{2}{The Observatories of the Carnegie Institution of Washington, 
813 Santa Barbara St., Pasadena, CA 91101; lho@ociw.edu}


\begin{abstract}
We have assembled a sample of 64 late-type spiral galaxies ($T$ types 6.0--9.0, corresponding to Hubble types Scd--Sm) with archival {\em Chandra} data. At a signal-to-noise (S/N) threshold of 3, we find 12 objects with X-ray point-source detections in close proximity with the optical or near-infrared position of the nucleus (median offset $\delta$ = 1\farcs 6), suggestive of possible low-luminosity active galactic nuclei (AGNs). Including measurements with 3 $>$ S/N $>$ 1.5, our detections increase to 18. These X-ray sources range in luminosity from $L_{\rm X}$(2--10 keV) = $10^{37.1}$ to $10^{39.6}$ ergs~s$^{-1}$. Considering possible contamination from low-mass X-ray binaries (LMXBs), we estimate that $\sim$5 detections are possible LMXBs instead of true AGNs, based on the probability of observing a LMXB in a nuclear star cluster typically found in these late-type spiral galaxies. Given the typical ages of nuclear star clusters, contamination by high-mass X-ray binaries is unlikely.  This AGN fraction is higher than that observed in optical surveys, indicating that active nuclei, and hence central black holes, are more common than previously suggested. The incidence of AGN activity in such late-type spiral galaxies also suggests that nuclear massive black holes can form and grow in galaxies with little or no evidence for bulges.  Follow-up multiwavelength observations will be necessary to confirm the true nature of these sources. 

\end{abstract}

\keywords{galaxies: active --- galaxies: nuclei --- galaxies: Seyfert --- galaxies: statistics --- X-rays: galaxies}


\section{INTRODUCTION}
There is now overwhelming evidence that the vast majority of massive early-type galaxies harbor a supermassive black hole (BH) at their centers, and that BHs play an important role during the assembly and evolution of their host galaxies (see reviews in \citealt{Ho04}). The existence of the $M_{\rm BH}-\sigma_\star$ correlation between BH mass and central stellar velocity dispersion links the growth of the BH with the formation of the host bulge \citep{Gebhardt00,FM00}, perhaps through feedback mechanisms associated with an active galactic nucleus (AGN). The detection of BHs can be achieved by several methods, including direct kinematics of surrounding gas and stars (e.g., \citealt{Kormendy04,Ghez05}), and optical spectroscopic identification of nuclear emission lines originating from an AGN (e.g., \citealt{Ho_2_95}).  The first method is limited to nearby, inactive or mildly active galaxies, however, while optical spectroscopic identification is generally limited to moderately bright nuclei that are not strongly confused by circumnuclear star formation or heavily obscured by dust.

An extensive spectroscopic survey of nearby galaxies has revealed that approximately $\sim 60$\% of galaxies of Hubble type E -- Sb contain an AGN \citep{Ho_5_97}, broadly consistent with the idea that BHs are ubiquitous in these systems \citep{Ho08}. For types later than Sc, where classical bulges disappear \citep{KK04}, only $\sim 15$\% of galaxies show optical AGN activity. This leads to the notion that classical bulges or ellipticals are necessary for BH formation. There are, however, a few striking counter examples. NGC\,4395, an essentially bulgeless Sm galaxy, exhibits all the hallmarks of a Seyfert 1 nucleus \citep{FS89,FHS93}, and it has a BH mass of $\sim 10^5 M_\odot$ \citep{FH03,Peterson05}. POX\,52 shares an almost identical optical spectrum to NGC\,4395, but is hosted in a spheroidal galaxy \citep{Barth04}\footnote{\cite{Barth04} called POX 52 a dwarf elliptical galaxy, but here we follow the nomenclature advocated by \cite{Kormendy08} and adopted in \cite{GHB08}.}, a visually similar but distinct morphological type compared to elliptical galaxies or classical bulges. More recently, an AGN was discovered in NGC\,3621, a late-type Sd galaxy, using infrared spectroscopic observations \citep{Satyapal07}. On the other hand, the nearby Scd galaxy M\,33 has a firm upper limit on the mass of any central dark object of $<1500 M_\odot$ \citep{Gebhardt01}, far below the typical $10^5 M_\odot$ BHs found in galaxies such as NGC\,4395. Thus BHs can indeed exist in bulgeless systems, but exactly how common such systems are remains unknown.

From a spectral analysis of SDSS data, \cite{GH07c} have found a population of low-mass BHs ($M_{\rm BH} \approx 10^5$--$10^6 M_\odot$) accreting at appreciable Eddington ratios. The host galaxies from the initial sample of 19 sources \citep{GH04}, using {\em HST} imaging, are found to be mid- to late-type spirals or spheroidals similar to POX\,52 \citep{GHB08}. None are as late-type as NGC\,4395, however, and so the incidence of BHs in very late-type systems remains uncertain.

For such systems, there is also the possibility that BHs are replaced by nuclear star clusters (NCs). Both \cite{Ferrarese06} and \cite{WH06} suggest that BHs in the $M_{\rm BH}-\sigma_\star$ relation are replaced by compact massive objects at low mass, with a generalized correlation $M_{\rm CMO} - \sigma_\star$. Such compact objects could include NCs instead of BHs, commonly found in late-type spirals \citep{Boker02,Boker04}. It is therefore not immediately obvious that such late-type spirals {\em should} contain BHs. Establishing a BH census for late-type spirals is important to determine if BHs are indeed formed in earlier-type galaxies only (ellipticals or early-type spirals), or if they also exist in very late-type galaxies.

One complication in late-type spirals is the abundance of gas and dust, with strong levels of star formation.  If these systems contain BHs, they are likely to be of relatively low mass, and hence the AGN, even when fully active, will be of quite low luminosity.  While the mere presence of an AGN indicates a BH, energetically weak or heavily embedded AGNs are extraordinarily difficult to observe, particularly at optical wavelengths where most of these observations have historically occurred.  Dust obscuration severely attenuates optical and ultraviolet flux, and central star formation can dominate the optical/infrared emission.

X-ray observations allow us to peer through the obscuring gas and dust column, since high column densities are required to suppress X-ray flux (particularly energetic X-rays). X-ray detections are therefore a useful indicator of AGN activity, and since the launch of the {\em Chandra X-ray Observatory}, the sensitivity with which to carry out such studies has been greatly improved. In particular, the Advanced CCD Imaging Spectrometer (ACIS; \citealt{Garmire03}) provides better than 0\farcs5 resolution with very low background contamination ($\sim 10^{-7}$ counts pix$^{-1}$ sec$^{-1}$ between 0.5 and 8 keV). This allows low flux limits to be achieved with very short (few ks) exposures, enabling surveys of non-X-ray selected samples \citep{Ho01}. \cite{Gallo07} have used X-ray detections successfully to identify BHs in low-mass bulges, and have begun to establish statistics on AGN X-ray activity as a function of host bulge mass. X-ray observations have also been useful in searching for nuclear activity in quiescent (i.e. optically faint) BHs residing in massive elliptical galaxies \citep{DiMatteo00,Loewenstein01,Pellegrini05,Soria06,WTH08}, and in star-forming early-type spiral galaxies \citep{TG07}.

In this paper we investigate the nuclear activity in a sample of late-type spiral galaxies using archival {\em Chandra} data, which provide a clearer method of detecting AGN activity in such star formation-dominated, potentially heavily obscured environments.  Systems with nuclear X-ray detections coincident with the optical nucleus are identified as AGN candidates, and after considering the possible contamination of our results from X-ray binaries, we estimate the prevalence of AGNs in late-type spiral galaxies. We discuss our sample selection in \S~\ref{sec:sample} and our results in \S~\ref{sec:results}. We conclude with a discussion of possible non-AGN sources and directions for future work in \S~\ref{sec:disc}.

\section{X-RAY SAMPLE AND DATA ANALYSIS}
\label{sec:sample}

All data were obtained from the {\em Chandra} Data Archive, with the full list summarized in Table~\ref{tab:log}. We initially selected objects from the Palomar spectroscopic survey of nearby galaxies \citep{Ho_3_97} with morphological $T$ types between 6.0 and 9.0 (representing Hubble types Scd to Sm; i.e. bulgeless or nearly bulgeless spiral galaxies), which were available in the archive. The Palomar survey contains the most sensitive and comprehensive statistics on local AGNs to date, but the reliability of optically derived statistics in late-type spirals is uncertain due to dust and star formation contamination. We can therefore complement the survey with X-ray data. Out of 63 objects in the Palomar survey with $T$ types between 6.0 and 9.0, 33 are present in the {\em Chandra} archives. The 33 Palomar objects in our sample have a median absolute magnitude of $M_B = -19.0$ (corrected for Galactic and internal extinction), and a median distance of 7.8 Mpc. 

If we do not restrict ourselves to Palomar objects, we can expand our sample significantly. From the master list of all public {\em Chandra} observations (as of May 2008), we searched for any galaxies in the HyperLeda\footnote{{\tt http://leda.univ-lyon1.fr}} database within 4\arcmin \ of the aimpoint (i.e. half a chip size for ACIS-S). After restricting our search to objects with $T$ type between 6.0 and 9.0 \citep{RC3} and within 30 Mpc \citep{Tully88}, our sample grows to 64 objects. These Palomar and non-Palomar subsamples are denoted by ``P'' and ``C,'' respectively, in Table~\ref{tab:log}. When multiple observations exist for a given object, we generally chose the longest ACIS-S exposure if possible, otherwise opting for the newest observation. In two cases ACIS-S images were not available. For M\,33, we chose the HRC-S observation, since the detector does not suffer from pileup as a result of the high count rate (see below for more discussion on pileup).

\begin{figure}
\epsscale{1.0}
\plotone{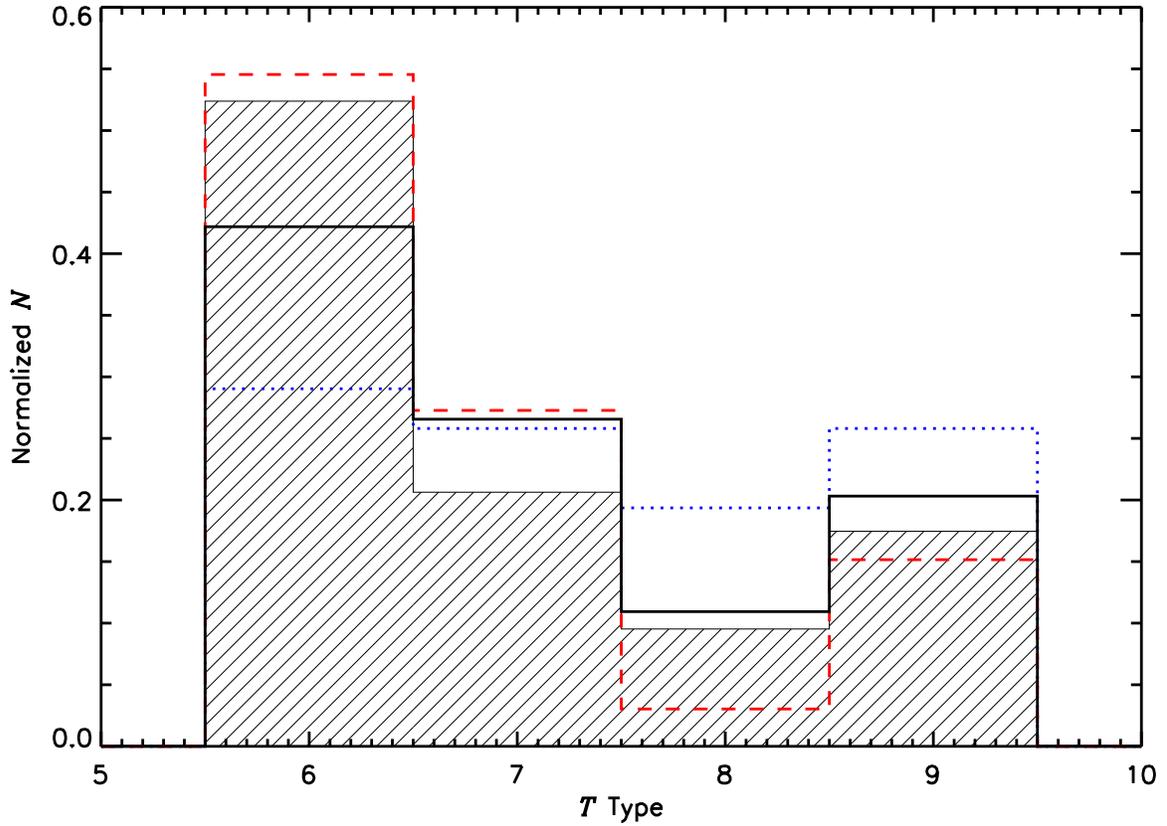}
\caption{Distributions of morphological $T$ types. The \cite{Ho_3_97} Palomar sample (with $T$ = 6.0--9.0) is shown as the shaded histogram. Our {\em Chandra} sample is shown as three unfilled histograms: the Palomar objects (P sample; red, dashed), the non-Palomar objects (C sample; blue, dotted), and all objects (black, solid) . All histograms have been normalized such that the total number over all $T$ types equals 1.}
\label{fig:ttypes}
\end{figure}

Because our sample is selected only on the basis of availability within the {\em Chandra} Data Archive, there is a potential for selection biases (based on, for instance, the selection criteria of the original proposals)---although these objects were not necessarily targeted for their nuclear properties. In Figure~\ref{fig:ttypes} we compare the distributions of $T$ type between our sample and the original sample of 63 Palomar objects from \cite{Ho_3_97}. The distributions agree reasonably well, although our sample is perhaps weighted slightly toward the most late-type spirals (i.e. $T$ = 9.0). 

With two exceptions, all data were obtained using the {\em Chandra} ACIS detector, in either the I- or S-array. The I-array consists of 4 CCDs arranged for imaging, while the S-array consists of 6 CCDs used for either imaging or grating spectroscopy. The mirror assembly has a point-spread function (PSF) with a full width at half maximum (FWHM) of 0\farcs5, equal to the ACIS pixel size. At a distance of 10 Mpc, this corresponds to a physical scale of $\sim$25 pc. Two objects were observed with the High Resolution Camera (HRC), which has 0\farcs4 pixels. 

We used type 2 event files output by standard pipeline processing. We used 0.2--8 keV counts in our analysis, limiting background contamination at high energies. All data were corrected for known astrometry offsets, following standard CIAO threads\footnote{\tt{http://cxc.harvard.edu/ciao/threads/}} (using CIAO version 3.3). We then used {\tt celldetect}, with default parameters, to automatically find all point sources within each image and determine the centroids. These detections have absolute astrometry accurate to 1\arcsec, although it can be marginally worse for off-axis objects with larger PSFs and few counts. From these centroids, we measured background-subtracted counts in a 2\arcsec\ radius aperture using the CIAO task {\tt dmextract}, with a background annulus ranging from 5\arcsec\ to 10\arcsec. In some cases, the background region was slightly altered to avoid nearby point sources.  For NGC\,3079 and 4945, the aperture is 1\arcsec\ to avoid nearby sources and strong background, and for NGC\,3184 the aperture is 3\arcsec\ since this object is far off-axis (and thus has a large PSF). A 2\arcsec\ aperture ensures that we measure virtually all of the point source counts, since the on-axis PSF of {\em Chandra} contains 95\% of the encircled energy with 1\arcsec.

Finally, we used Two Micron All-Sky Survey (2MASS; \citealt{2MASS06}) positions (obtained from the NASA/IPAC Extragalactic Database, NED), with typical $1\,\sigma$ uncertainties of $\sim$0\farcs5 \ for the photometric center of {\em extended} sources \citep{Jarrett00}, to determine whether X-ray point-source detections were coincident with the nucleus. While the nominal accuracy for 2MASS is often quoted as 0\farcs1, this only applies to point sources. The distribution of offsets between the nearest X-ray detection and photometric center from 2MASS is highly bimodal, and we chose to consider objects belonging to the smaller-offset group as AGN candidates. Our formal criterion requires that the 2MASS positions and X-ray detections lie within 5\arcsec \ of each other. In practice almost all candidates have better than 2\farcs5 agreement, while nearly all non-detections have no X-ray point sources within 10\arcsec \ of the 2MASS position. (NGC\,4625 has an X-ray source within 6\arcsec \ of the optical nucleus, while NGC\,3395, NGC\,3556, NGC\,4038, and NGC\,4631 have an X-ray source within 10\arcsec \ of the optical nucleus.) The minimum X-ray signal-to-noise (S/N) threshold accepted was 1.5, although in practice only three sources were detected with such low S/N. We detected 18 X-ray point sources in our sample of 64 objects; 12 sources with S/N $>$ 3, three sources with 3 $>$ S/N $>$ 2, and three sources with 2 $>$ S/N $>$ 1.5.

For all AGN candidates except IC\,342, M\,33, NGC\,55, and NGC\,4395, we estimate the total X-ray luminosity similarly to \cite{Ho01}, assuming a photon index of $\Gamma =1.8$, typical of most low-luminosity AGNs (see \citealt{Ho08}), and an average Galactic interstellar medium column density of $N_{\rm H} = 2 \times 10^{20}$ cm$^{-2}$.  This gives $L_{\rm X}$(2--10 keV) $ = 3.6 \times 10^{37}$ ergs s$^{-1}$ (ACIS counts ks$^{-1}$) ($D$/10 Mpc)$^2$. This is only valid, however, for ACIS-S data. For IC\,342 and NGC\,55 (i.e. HRC-I and ACIS-I data), we use the PIMMS tool\footnote{\texttt{http://cxc.harvard.edu/toolkit/pimms.jsp}} (version 3.9d) with the same assumptions to obtain the 2--10 keV luminosity. The luminosities would increase by a factor of 1.8 and 6.7 for intrinsic column densities of obscuring gas and dust of $2 \times 10^{21}$ and $2 \times 10^{22}$ cm$^{-2}$, respectively. For M\,33 (HRC-S data), we again use PIMMS but assume $N_{\rm H} = 2 \times 10^{21}$ cm$^{-2}$ \citep{Takano94,Parmar01,DR02,LaParola03}. NGC\,4395 is a known AGN with strong absorption ($N_{\rm H} \approx 2 \times 10^{22}$ cm$^{-2}$; \citealt{Iwasawa00, Moran05}), and so we use this value instead to estimate $L_{\rm X}$. None of our ACIS objects suffer from significant pileup (multiple detections of lower energy photons within a single frame time that is registered as a single event at higher energy, possibly even rejected as a cosmic ray hit). NGC\,4395 has the highest count rate of $7.5\times10^{-2}$ counts s$^{-1}$, which corresponds to a pileup of only 2\% (calculated using PIMMS), roughly equal to the measured uncertainty in total counts. 

We obtained an upper limit for those sources without an X-ray point-source detection. We used a circular aperture with a 2\arcsec \ radius, centered at the optical position of the nucleus. We measured the background counts with its uncertainty, and determined\footnote{See the online \texttt{celldetect} documentation for a detailed description of how the S/N ratio is calculated.} the counts necessary for a theoretical source to have a S/N equal to 1.5 and thus be detected by {\tt celldetect}. These counts were then converted to a luminosity as described above. In some cases, because of the very low exposure times, no counts were detected in this aperture size. In these cases, we used larger apertures and scaled down to a 2\arcsec \ aperture, up to a 12\arcsec \ aperture in the most extreme cases. Fields were sparse enough that no contamination from nearby X-ray point sources was present. For objects with a detected X-ray point source, the limiting luminosity was similarly found using the background measurements, scaled to a 2\arcsec\ aperture.

As a proxy for spectral shape, we compute X-ray colors for those objects with detected X-ray sources, with counts measured in three bands: soft ($S$, 0.3--1 keV), medium ($M$, 1--2 keV), and hard ($H$, 2--8 keV). The soft and hard colors are then defined to be $H_1 \equiv (M-S)/T$ and $H_2 \equiv (H-M)/T$, respectively, where $T$ is the total counts in all three bands. This was only achievable with ACIS data, which have accurate registration of incoming photon energy, whereas the HRC instrument has a spectral resolution of only $E/\Delta E \approx 1$. 

Four objects in our sample have sufficient counts (i.e. $\gtrsim 200$) to do a preliminary spectral analysis: NGC\,4039, NGC\,4395, NGC\,4490, and NGC\,6946. NGC\,4945 has most of its counts in a strong Fe K$\alpha$ feature at 6.4 keV, resulting in few continuum photons. To accomplish this analysis, we used the CIAO task {\tt psextract} to extract the X-ray spectra along with the associated response files. The aperture and background used are the same as those for the X-ray photometry. Bins are selected to have a minimum of 20 counts. Using {\tt XSPEC} \citep{Arnaud96}, we then fit simple models to the data, ignoring bins with energies below 0.3 keV and above 5 keV (where there are few counts). In the case of NGC\,4395, a strong source, we ignored bins with energies below 1 kev (there is a soft excess at low energies; see also \citealt{Moran05}) and above 10 keV. We tested various models, including an absorbed power law, a blackbody, and an absorbed power law + blackbody. 

\section{RESULTS}
\label{sec:results}

\begin{figure}
\epsscale{0.9}
\plotone{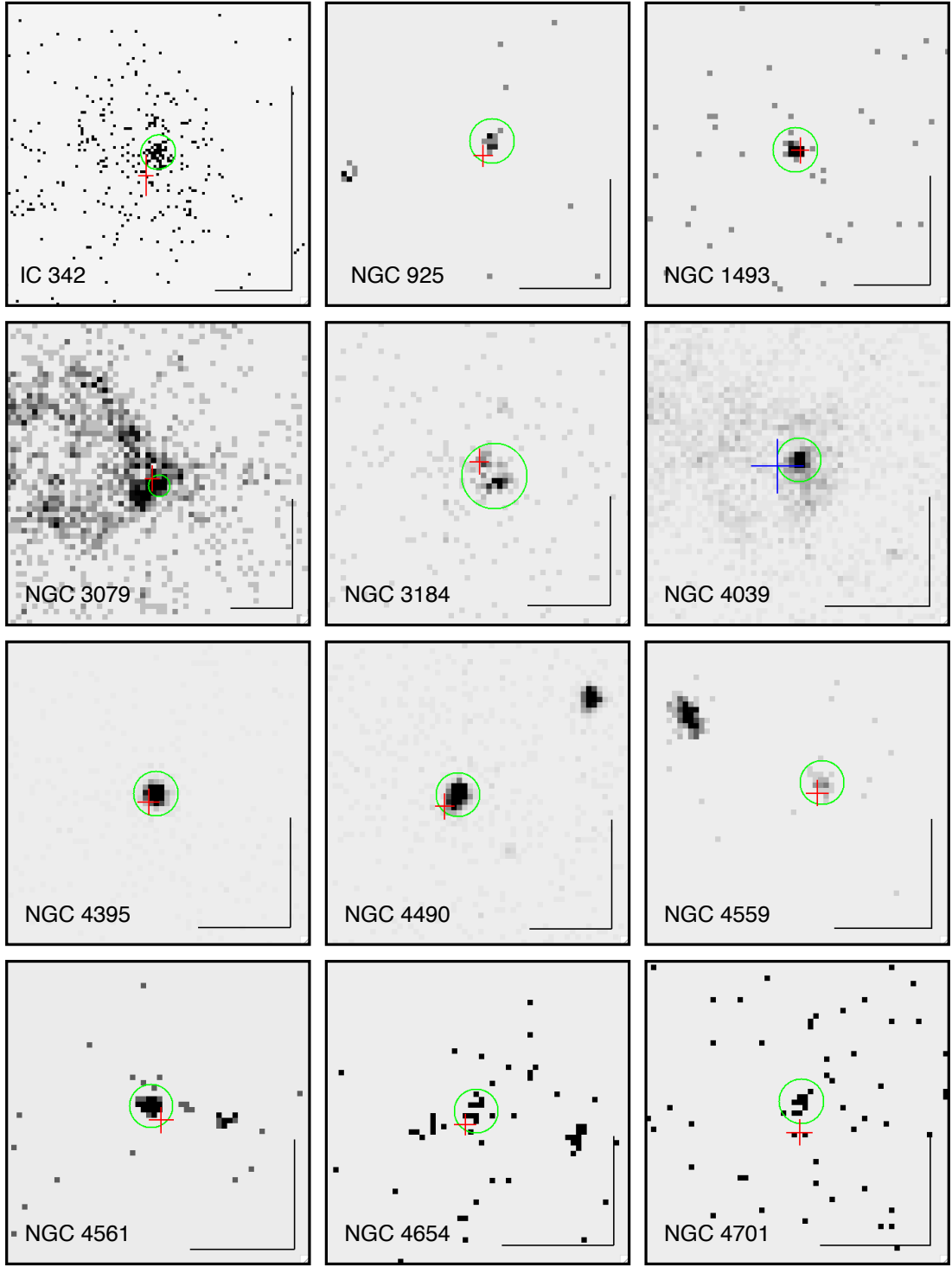}
\caption{Images in the 0.2--8 keV band of all AGN candidates from Table~\ref{tab:agn}. Green circles identify point-source detections in the image coincident with the optical nucleus. Crosses locate the position of the nucleus as found from 2MASS (red) or other optical sources (blue). See Table~\ref{tab:log} for more details. The size of the cross represents the 95\% confidence uncertainty in the near-infrared/optical position. The pixel scale varies slightly from object to object, and in some cases is non-square. A $10\arcsec \times 10\arcsec$ guide is shown on the lower right of each image.}
\label{fig:images}
\end{figure}

Out of our sample of 64 objects, we detect 18 sources coincident with the optical position of the nucleus. We consider these as AGN candidates (except for M\,33, which we discuss in Section~\ref{sec:disc}). Our full sample is summarized in Table~\ref{tab:log}, while the AGN candidates are summarized in Table~\ref{tab:agn}, along with their measured X-ray colors. Their images are presented in Figure~\ref{fig:images}. Of the 17 candidates, three objects are known AGNs from optical diagnostics (NGC\,3079, NGC\,4395, and NGC\,4945, although they are weak sources), and one more is a known low-ionization  nuclear emission-line region (LINER; NGC\,6503). NGC\,4654, NGC\,4701 and NGC\,5457 were detected above a S/N of 1.5. NGC\,925, NGC\,4713, NGC\,6503 were detected above a S/N of 2. The rest of the AGN candidates were detected with a S/N above 3. NGC\,45 has a possible nuclear X-ray source coincident with the optical position, but it was not detected above a S/N of 1.5, and thus is not included in our candidate list.

\begin{figure}
\epsscale{1.0}
\plotone{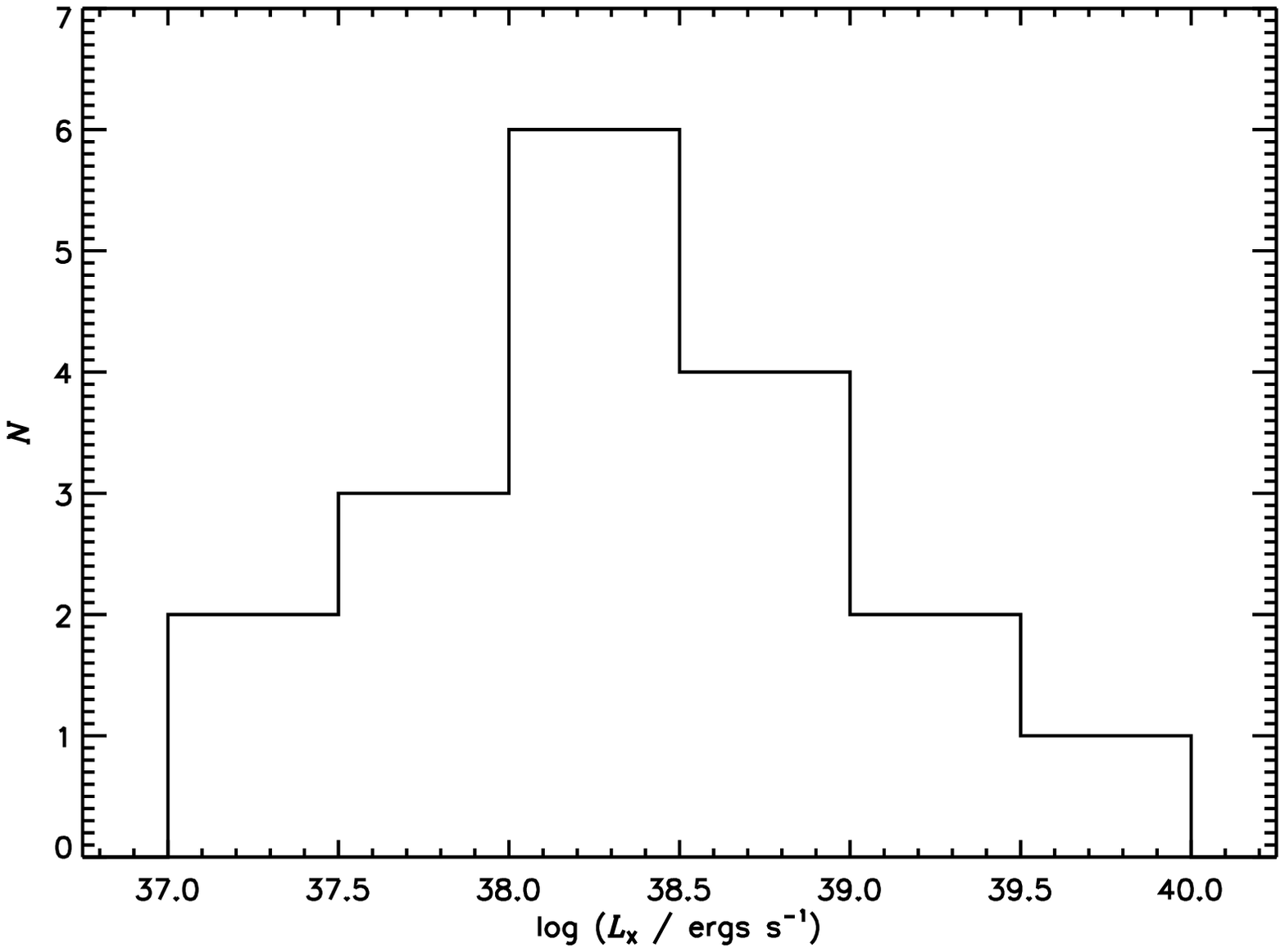}
\caption{Distribution of 2--10 keV X-ray luminostity for the AGN candidates, as listed in Table~\ref{tab:agn}. $L_{\rm X}$ is determined using distances from \cite{Tully88} (unless otherwise noted in Table~\ref{tab:log}) and assuming a power law with a photon index of $\Gamma =1.8$ and Galactic column density of $N_{\rm H} = 2 \times 10^{20}$ cm$^{-2}$ \citep{Ho01}, except for M\,33, where we assume $N_{\rm H} \approx 2 \times 10^{21}$ cm$^{-2}$ \citep{Takano94,Parmar01,DR02,LaParola03}, and NGC\,4395, where we assume $N_{\rm H} \approx 2 \times 10^{22}$ cm$^{-2}$ \citep{Iwasawa00, Moran05}.}
\label{fig:lx}
\end{figure}

The 17 AGN candidates have 2--10 keV luminosities ranging from $L_{\rm X} = 10^{37.1}$ to $10^{39.6}$ ergs s$^{-1}$. The distribution of $L_{\rm X}$ is shown in Figure~\ref{fig:lx}. The three known AGNs (NGC\,3079, NGC\,4395, and NGC\,4945) have  $L_{\rm X} = 10^{38.2}$, $10^{39.6}$, and $10^{37.7}$  ergs s$^{-1}$, respectively. The median offset between the X-ray detection and the optical/near-infrared position is $\delta$ = 1\farcs 6. For those objects with enough counts, our X-ray spectral fits are summarized in Table~\ref{tab:spectra} and Figure~\ref{fig:spectra}, and are discussed below.

For the AGN candidates with available measurements of $\sigma_\star$ in the literature, we estimate a possible BH mass using the $M_{\rm BH}-\sigma_\star$ relation of \cite{Tremaine02}. We also estimate the corresponding X-ray luminosity to Eddington luminosity ratio. These are presented in Table~\ref{tab:agn}.  

\section{DISCUSSION}
\label{sec:disc}

The evidence for the majority of these candidates being true AGNs is circumstantial. For most of these systems, the physical scale of an X-ray point source in {\em Chandra} data is of order tens of parsecs, making it extremely unlikely that the vast majority of detections are serendipitous field X-ray binaries along the line of sight to or near the nucleus. Furthermore, the probability of detecting a background/foreground X-ray source within the {\em Chandra} PSF is extremely low, of order $10^{-7}$ \citep{Gallo07}. With a few exceptions, the positional agreement between the X-ray source and the 2MASS position of the nucleus is excellent (less than 2\arcsec, compared to the {\em Chandra} and 2MASS positional uncertainties of $\sim$1\arcsec). NGC\,4945 has a larger offset ($\delta \approx$ 3\farcs2), but the presence of bright diffuse X-ray emission makes point-source detection more challenging, and once again this object is a known Seyfert 2. 

Many of these galaxies have \hii \ nuclei as identified in the Palomar survey, but most of the optical observations have slit apertures of $2 \arcsec \times 4 \arcsec$ \citep{Ho_2_95}, much larger than the {\em Chandra} PSF. The Palomar data are therefore probing a much larger region than the X-ray point sources, and we must be careful when making comparisons. As these are gas- and dust-rich systems, there is clearly some level of star formation in the circumnuclear region. If we consider, for example, typical \ha \ luminosities emitted by such low-luminosity AGNs, they are significantly lower than the measured \ha \ luminosities in the Palomar data. Assuming the bolometric, X-ray, and \ha \ luminosities of AGNs are related by $L_{\rm bol} = 16 L_{\rm X} = 220 L_{\rm H\alpha}$ \citep{Ho08}, a range $L_{\rm X} = 10^{37}$ -- $10^{40}$ ergs s$^{-1}$ corresponds to $L_{\rm H\alpha} = 7 \times 10^{35}$ -- $7 \times 10^{38}$ ergs s$^{-1}$. Palomar objects in our sample with \hii \ nuclei have $L_{\rm H\alpha} = 10^{37.5}$ -- $10^{38.8}$ ergs s$^{-1}$, which can explain why such low-luminosity AGNs, if present, were missed in the Palomar data.

The galaxies in our sample are low-mass, late-type spirals, and thus likely harbor a NC at the center of the gravitational potential well. \cite{Boker02,Boker04} found that $\sim$75\% of late-type spirals ($T$ = 6--9) host a compact, photometrically distinct source located near the photometric center of the galaxy, with the surface brightness profile deviating from a pure disk profile. These NCs have mean effective radii of $r_{\rm eff} = 3.5$ pc and $I$-band luminosities of $L_I=10^{6.2} L_\odot$. \cite{Rossa06} found that for NCs in galaxies within a similar range in $T$ type, the average cluster mass is $M=10^{6.25} M_\odot$. The increased stellar density at the center therefore implies an increased chance of detecting, in particular, a low-mass X-ray binary (LMXB) as a nuclear X-ray point source, confusing our results. M\,33, for example, likely has no central massive BH \citep{Gebhardt01}, but does possess an X-ray source coincident with the optical nucleus, with properties suggestive of an X-ray binary \citep{Dubus04}. This is why we do not included M\,33 as part of our AGN candidates. NGC\,2403 also hosts an X-ray binary in its known nuclear star cluster \citep{Yukita07}, although the X-ray luminosity is not high enough to be confused with an AGN (in this work, no detection was found above the limiting luminosity of $10^{36.1}$ ergs s$^{-1}$). X-ray binaries must therefore be carefully considered when attempting to detect low-luminosity AGNs. Although NCs also occur frequently in more massive early-type spiral galaxies \citep{Carollo97,Carollo98} and intermediate-luminosity elliptical galaxies \citep{Cote06}, LMXB contamination is not important in these cases because at these larger BH masses $L_{\rm X} > 10^{40}$ ergs s$^{-1}$ due to nuclear activity, much greater than typical LMXB X-ray luminosities.

An X-ray detection in a NC is much more likely to be a LMXB rather than a high-mass X-ray binary (HMXB) because the mean age of NCs in late-type hosts is $\sim 10^{8 {\rm -} 9}$ yrs \citep{Rossa06}, older than typical high-mass stars in HMXBs. Although many late-type galaxies have \hii \ nuclei \citep{Ho_3_97}, the slit aperture used, in most cases, is significantly larger than the {\em Chandra} PSF. Given that nearly all galaxies in the survey exhibit some emission lines in their nuclear spectra, and that late-type galaxies in particular are gas and dust rich with strong levels of star formation, the prevalence of \hii \ nuclei is not surprising. This does not necessarily imply, however, that the NCs have ongoing star formation (and therefore HMXBs are more likely), as evidenced by the inferred NC ages from \cite{Rossa06}.

For our purposes, we will approximate the NCs as globular clusters (GCs) in order to estimate the incidence of LMXBs. Although the environments of both types of clusters are widely different, NCs and GCs share remarkably similar structural and dynamical properties. Viewed in the fundamental plane (relating effective radius, surface density, and velocity dispersion), NCs occupy a region very distinct from classical bulges and elliptical galaxies, and fall on the high-mass extension of the GC sequence \citep{Walcher05}. With some important exceptions (see below), approximating NCs as GCs is therefore well justified. 

As in \cite{Gallo07}, we can estimate the expected number of LMXBs brighter than $3.2 \times 10^{38}$ ergs s$^{-1}$ for a GC (and thus NC) of given mass, size, and $g-z$ color, using an expression derived by \cite{Sivakoff07}:

\begin{equation}
n_{\rm X} = 8 \times 10^{-2} \left( \frac{M}{10^6 M_\odot} \right)^{1.237} 10^{0.9(g-z)} \left( \frac{r_{\rm h,corr}}{1 {\rm \ pc}} \right)^{-2.22}.
\end{equation}

\noindent
The quantity $r_{\rm h,corr}$ specifies the corrected half-mass radius, where $r_{\rm h,corr} = r_{\rm h} \times 10^{0.17[(g-z)-1.2]}$. Since this depends rather weakly on color, we can approximate $r_{\rm h,corr} \approx r_e = 3.5$ pc as a mean value for NCs in late-type spirals. \cite{Rossa06} find a mean $B-V$ color for such NCs of order $\sim 0.5$--$0.6$ mag. In order to get $g-z$ colors, we use approximate color transformations from \cite{Fukugita95} by approximating the stellar population in a NC as that between an Scd and an Sbc galaxy, which have $B-V$ colors similar to those of NCs. This gives $g-z$ colors ranging from 1.05 to 1.17 mag; we adopt $g-z = 1.1$ mag. Therefore the mean number of LMXBs we expect is $n_{\rm X} \approx 0.1$. This estimate is specifically for sources brighter than $\sim 10^{38}$ ergs s$^{-1}$, which corresponds to the completeness limit of \cite{Sivakoff07}. The limiting luminosity for our X-ray detections varies from $L_{\rm X} = 10^{36.2}$ to $10^{38.1}$ ergs s$^{-1}$. For objects in the \citeauthor{Sivakoff07} sample with a lower limiting luminosity ($\sim 10^{37}$ ergs s$^{-1}$ or lower), however, they detected only a factor of a few more sources. 

An important caveat to this estimate is the question of stellar age. GCs generally have old stellar populations, formed via a single burst of star formation, whereas NCs have repeated star formation episodes since the reservoir of gas can be replenished. This leads to, for a given cluster mass, a younger mean stellar age for NCs compared to GCs, and thus bluer colors \citep{Rossa06,Walcher06}. As a result, the expected number of LMXBs estimated above is an upper limit, since it was derived for GCs, and the true number of LMXBs in NCs is likely lower. This is due to the $\sim 1$ Gyr timescale necessary for the stellar companion to come into Roche lobe contact with its BH companion (and thus turn on the LMXB), either by stellar evolution or via orbital decay \citep{WG98}.

Using $n_{\rm X} \approx 0.1$ as our estimate, this would imply only $\sim 5$ X-ray detections from LMXBs in a sample of 64 late-type spirals (assuming 75\% have NCs with mean properties), of which M\,33 is a typical example. This is a factor of 3--4 lower than our observed detection rate. The observed X-ray luminosities in our sample are similar to the known AGNs, making our other detections reasonable AGN candidates. This implies that NCs and BHs need not be necessarily mutually exclusive, if the majority of our AGN candidates are true AGNs. Indeed, \cite{Seth08} find that over all Hubble types, roughly 10\% of all galaxies hosting a NC also host an AGN, based on optical spectroscopy, with a fraction that increases strongly with galaxy and NC mass. 

It is important to note that these detections are AGN {\em candidates}, and are worthy of further detailed observations to properly identify these sources. \cite{Ghosh08} provide some evidence from a multiwavelength analysis that NGC\,3184, NGC\,4713, and NGC\,5457 are true AGNs. Since most of our candidates have very few counts, long-integration X-ray spectroscopy will be useful in differentiating true AGNs from LMXBs, particularly at very soft and hard energies. As described above, AGNs have X-ray spectra composed of an absorbed power law (with $\Gamma \approx 1.8$), whereas X-ray binaries have spectra that can be separated into three states: a high/soft state with a $\sim$1 keV thermal component and a weak nonthermal tail at high ($>$ 10 keV) energies; a very high state with a steep power law ($\Gamma \approx 2.5$) combined with a $\sim$1 keV thermal blackbody, which can constitute a significant amount of the total flux; and a low/hard state with a single power law ($1.4 < \Gamma < 2.1$) \citep{RM06}. Of these three states, only the low/hard state could be confused with an AGN, but X-ray binaries in these states typically have extraordinarily low luminosities ($L_{\rm X} = 10^{30.5}$ -- $10^{33.5}$ ergs s$^{-1}$), which would distinguish them from true AGNs. 

\begin{figure}
\epsscale{1.0}
\plotone{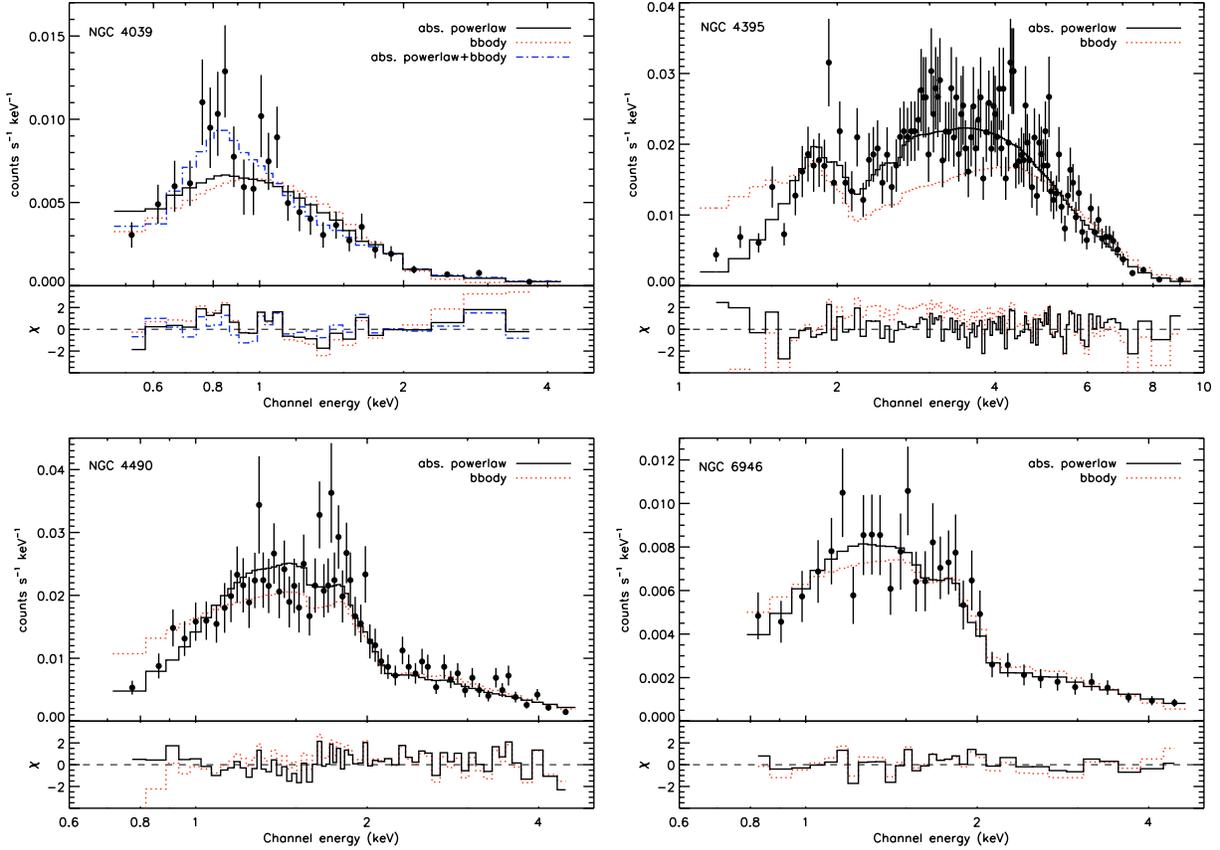}
\caption{Extracted X-ray spectra for objects with $\gtrsim 200$ counts (except for NGC\,4945; see text). Energy bins are chosen to have a minimum of 20 counts. Also plotted are various fitted models, including an absorbed power law (solid black line), a thermal blackbody (dotted red line), and an absorbed power law + blackbody (dash-dotted blue line). Model fit parameters are given in Table~\ref{tab:spectra}. The bottom panel in each plot shows the residuals for each model normalized by the $1\sigma$ uncertainty in the measurement.}
\label{fig:spectra}
\end{figure}

Our spectral fits (Table~\ref{tab:spectra} and Figure~\ref{fig:spectra}) are insufficient to definitively rule out one model over another, though they remain very suggestive. For NGC\,4039, a pure thermal blackbody is likely ruled out, whereas an absorbed power law is plausible, and an absorbed power law + blackbody model (indicative of a very high state LMXB) is consistent with the data. Given that the reduced $\chi^2$ of the latter model is $<$1, however, we may be overfitting the data, and so we must be cautious. Although it may appear as though this is more likely an X-ray binary (of extremely high luminosity $>10^{39}$ ergs s$^{-1}$), longer-integration data are needed to distinguish between the two models. 

NGC\,4490 and NGC\,6946 are both well fit by simple absorbed power law models ($\chi^2 \approx 1$), consistent with the AGN interpretation. In the case of NGC\,4490, the unabsorbed blackbody model results in a poorer fit, particularly at low energies, while for NGC\,6946, both models are consistent with the data. Unfortunately, the data are not of high enough quality to differentiate between models with more parameters, such as an absorbed power law + blackbody, since $\chi^2 \lesssim 1$ in all cases. Once again, these results underscore the need for higher S/N data if we are to use a $\chi^2$ statistic to differentiate between various models.. As a check, we also analyzed NGC\,4395, a known Seyfert 1, which is unsurprisingly well fit by an absorbed power law, whose parameters are consistent with those given by \cite{Moran05}.

\begin{figure}
\epsscale{1.0}
\plotone{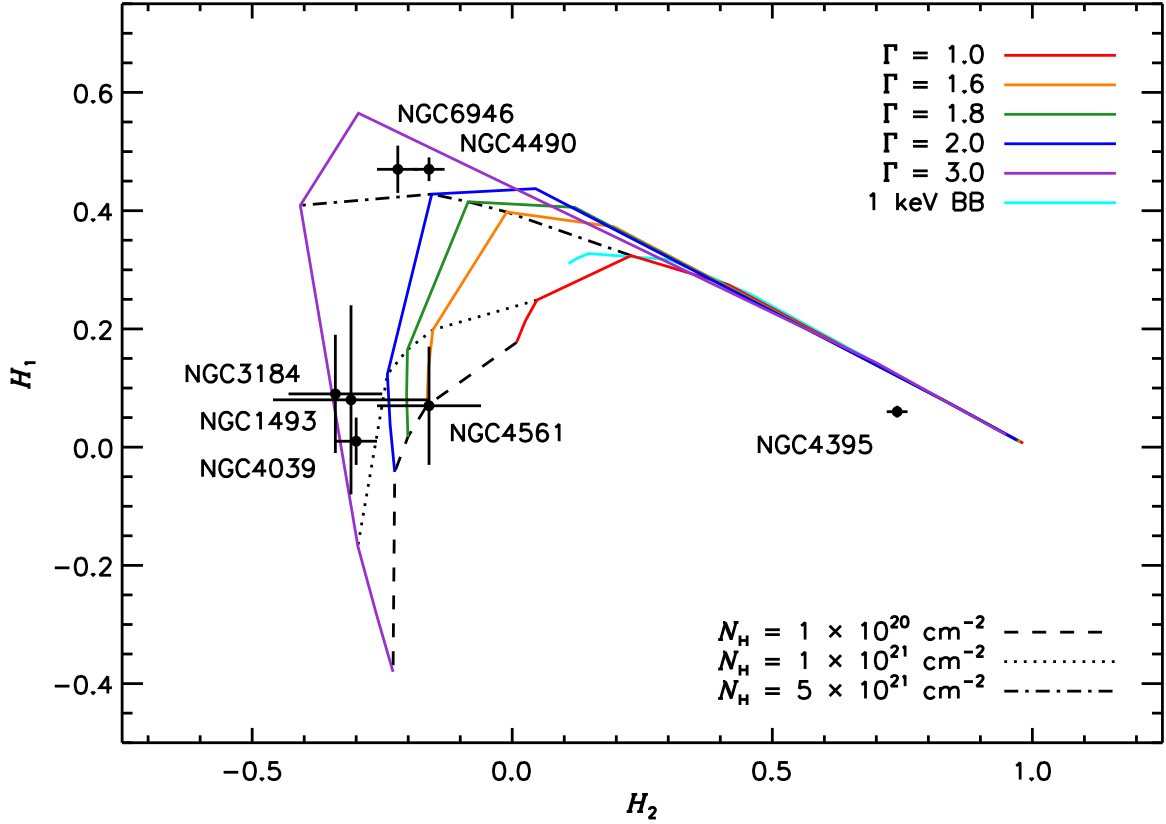}
\caption{X-ray colors of AGN candidates with sufficient counts, from Table~\ref{tab:agn}. Colors are defined such that $H_1 \equiv (M-S)/T$ and $H_2 \equiv (H-M)/T$, where $S$, $M$, $H$, and $T$ are defined to be the soft (0.3--1 keV), medium (1--2 keV), hard (2--8 keV), and total (0.3--8 keV) counts, respectively. Also plotted are the expected colors for absorbed power-law models (observed with ACIS-S) with photon indices of 1.0 (red), 1.6 (orange), 1.8 (green), 2.0 (blue), and 3.0 (purple), as well as a 1 keV blackbody (cyan). The column density $N_{\rm H}$ for each model ranges from  $1 \times 10^{20}$ cm$^{-2}$ to $1 \times 10^{23}$ cm$^{-2}$. Lines of constant column density for the power-law models are overplotted for clarity, in particular $N_{\rm H} = 1 \times 10^{20}$ cm$^{-2}$ (dashed), $N_{\rm H} = 1 \times 10^{21}$ cm$^{-2}$ (dotted), and $N_{\rm H} = 5 \times 10^{21}$ cm$^{-2}$ (dash-dotted). All AGN candidates with X-ray colors are consistent with absorbed power-law models, and inconsistent with a 1 keV blackbody, typical of an X-ray binary in the ``thermal'' state \citep{RM06}. Large uncertainties make it difficult to determine the photon indices and column densities precisely. NGC\,4395 is a known AGN with strong absorption ($N_{\rm H} \approx 2 \times 10^{22}$ cm$^{-2}$; \citealt{Iwasawa00, Moran05}).}
\label{fig:colors}
\end{figure}

In Figure~\ref{fig:colors} we show the X-ray colors of those AGN candidates with sufficient counts. We also include various absorbed power-law models and a 1 keV blackbody (representative of the high/soft state from \citealt{RM06}), computed for observations with ACIS-S using {\tt XSPEC}. All AGN candidates are consistent with absorbed power-law models, and inconsistent with the 1 keV blackbody. Although the uncertainties are large, it appears as though most of these candidates exhibit softer X-ray spectra than is typically assumed for low-luminosity AGNs. Low-mass AGNs can exhibit a range of X-ray spectra, however, with $\Gamma \approx$ 1.0 to 3.0 \citep{GH07a}. Clearly, detailed X-ray spectroscopy is needed to further characterize these systems. 

Once AGNs are positively identified, determining a BH mass estimate will be challenging. Dynamical measurements are currently not feasible in these objects. Reverberation mapping \citep{Peterson07} or virial estimates \citep{GH05} based on the strength of optical emission lines (if detected) will likely require excellent seeing conditions with a narrow slit or adaptive optics. As these are all nearly bulgeless systems, using the bulge velocity dispersion $\sigma_\star$ or the bulge luminosity $L_B$ \citep{MH03} to determine $M_{\rm BH}$ is challenging and highly uncertain. NCs with BHs, however, fall on the $M_{\rm BH}-\sigma_\star$ relation---for example NGC\,4395 \citep{FH03,Peterson05} and G1 \citep{GRH02,GRH05}---consistent with the extrapolation of $M_{\rm BH}-\sigma_\star$ to lower masses \citep{BGH05}. The velocity dispersions of NCs \citep{Walcher05} could therefore be used to estimate the BH mass. BH masses can also be estimated from X-ray variability timescales derived from complete power density spectra (\citealt{McHardy06}), obtained over long monitoring campaigns. Alternatively, the BH mass can be estimated from the excess variance measured in shorter X-ray data sets \citep{NPC04,GNC08}.

Our detection rate suggests an AGN fraction in late-type spiral galaxies of $\sim$20\%--25\% (depending on the LMXB contamination), higher than that observed by \cite{Ho_5_97}. (We should caution, however, that several detections are at low S/N. Using only S/N$>$3 detections, our AGN fraction becomes $\sim$17\%.) This would imply that we are indeed missing many low-mass AGNs in optical surveys, and that BHs in such systems are more prevalent than previously suggested. Optical surveys of low-mass AGNs can only find systems with high accretion rates, such that their bolometric luminosity $L_{\rm bol}$ approaches a substantial fraction of the Eddington luminosity $L_{\rm Edd} \equiv 1.26 \times 10^{38} (M_{\rm BH} / M_\odot)$ ergs s$^{-1}$ \citep{GH07b}. \cite{GH07c} found a median $L_{\rm bol}/L_{\rm Edd} = 0.4$ and a minimum $L_{\rm bol} / L_{\rm Edd} \approx 10^{-2}$ in a sample of 174 optically selected low-mass AGNs. In a pilot study drawn from this sample, \cite{GH07a} measured $L_{\rm X}$(0.5--2 keV) $\approx 10^{41}$--$10^{43}$ ergs s$^{-1}$, much higher than any of our measurements. Most of our candidates with available $M_{\rm BH}$ estimates exhibit $L_{\rm bol}/L_{\rm Edd}$ ratios orders of magnitude lower than those of \citeauthor{GH07c}, assuming a correction factor of 16 to convert from $L_{\rm X}$ to $L_{\rm bol}$ \citep{Ho08}. The one exception is NGC\,4395, a known Seyfert nucleus, with $L_{\rm bol} / L_{\rm Edd} \approx 10^{-2.1}$, consistent with the value of $2 \times 10^{-3}$ to $2 \times 10^{-2}$ found by \cite{FH03}. (In comparison, POX\,52 has $L_{\rm bol}/L_{\rm Edd} \approx 1$; \citealt{Barth04}.) Our estimated Eddington ratios are similar, however, to those seen in massive BHs in earlier-type hosts with low-luminosity AGNs \citep{Ho08}, suggesting that we are detecting the low-mass analogs of weakly accreting BHs. X-ray surveys therefore allow us to probe much fainter down the luminosity function of low-mass AGNs, detecting weakly accreting systems that would be completely swamped by galactic emission at optical wavelengths, and obtain a much higher detection rate. The \citeauthor{GH07c} sample, for instance, identified only 174 low-mass AGNs out of 8435 optically identified AGNs (from a parent sample of $\sim$600,000 galaxies). 

Given the low estimated accretion rates of our AGN candidates, radio observations can help to clarify the nature of these X-ray sources. If these are indeed low-mass and low-accretion-rate systems, we should expect them to have a radio signature, much like the more massive low-luminosity BHs \citep{Ho08}. Extrapolating from higher masses, we would expect 2 cm radio fluxes of order 10 to 100 $\mu$Jy \citep{Nagar02}. If these X-ray sources are X-ray binaries instead, however, they would be accreting at Eddington or super-Eddington rates to achieve such high luminosities. X-ray binaries in such states do no have a radio counterpart \citep{RM06}. 

\cite{Satyapal08} have conducted a similar survey of AGNs in the infrared, searching for emission-line signatures of AGN activity (in particular, \nev \ emission; see \citealt{AS08}) in late-type galaxies that are optically quiescent. This should identify weakly accreting systems similar to our candidates. In a sample of 32 objects, nine of which overlap with our {\em Chandra} sample, they find seven AGN candidates, a similar detection fraction. Interestingly, of the nine overlap objects, six (IC\,342, NGC\,925, NGC\,3184, NGC\,4490, NGC\,4559, and NGC\,6946) are identified in this paper as AGN candidates, but not by \citeauthor{Satyapal08} It is possible that the weakly active nuclei picked out by our X-ray search method are too faint to emit much \nev \ emission. We can estimate the total \nev \ luminosity $L_{\rm NeV}$ for our sample assuming the bolometric luminosity $L_{\rm bol} = 16 L_{\rm X}$ and $L_{\rm bol} = 9 L_{5100}$ \citep{Ho08}, where $L_{5100}$ is the continuum luminosity at 5100 \AA, coupled with the correlation between $L_{5100}$ and $L_{\rm NeV}$ found by \cite{Dasyra08}. For our range in X-ray luminosities ($L_{\rm X} = 10^{37}$ -- $10^{40}$ ergs s$^{-1}$), this corresponds to \nev \ luminosities of $L_{\rm NeV} = 10^{35}$ -- $7 \times 10^{37}$ ergs s$^{-1}$. These luminosities are orders of magnitude fainter than those measured by \cite{Satyapal08}. Even NGC\,3184, which appears to be a {\em bona fide} AGN \citep{Ghosh08}, was not detected. The \citeauthor{Satyapal08} detections of \nev \ are in excess of the \citeauthor{Dasyra08} relation, however, thus it is possible these relations do not hold for very late-type spiral galaxies. 

Another factor that might be relevant is the ionization parameter; since \nev\ is a high-ionization line, its strength decreases dramatically when the ionization parameter decreases \citep{AS08}.  By analogy with the local population of LINERs and Seyferts \citep{Ho08}, the very low Eddington ratios of our low-mass systems should correspond to low ionization parameters, and thus plausibly weak \nev\ emission.   Finally, we note that NGC\,3556 is identified by \citeauthor{Satyapal08} as an AGN candidate but has no X-ray detection in this work. \cite{WCI03} identify an X-ray source as a candidate for the galactic nucleus in NGC\,3556 (which they number source 35); we chose not to include this source in our candidate list as the offset from the optical nucleus is 10\arcsec. More detailed study is clearly warranted, and it will likely take the combined efforts of large multiwavelength surveys to find such low-mass and low-luminosity AGNs, and to properly assess the global AGN fraction in these late-type systems.

\acknowledgements
L.-B.D. would like to thank Eliot Quataert for the many helpful and insightful discussions, and Jenny Greene for advice and guidance. We also thank an anonymous referee for careful, helpful comments that greatly improved this manuscript. Support for this work was provided by NASA through Chandra Award Number 06700184 issued by the Chandra X-ray Observatory Center, which is operated by the Smithsonian Astrophysical Observatory for and on behalf of NASA under contract NAS8-03060, and through grant HST-GO-11130.01 from the Space Telescope Science Institute, which is operated by the Association of Universities for Research in Astronomy, Inc., for NASA, under contract NAS5-26555.  This research has made use of the NASA/IPAC Extragalactic Database (NED) which is operated by the Jet Propulsion Laboratory, California Institute of Technology, under contract with the National Aeronautics and Space Administration. We also acknowledge the usage of the HyperLeda database ({\tt http://leda.univ-lyon1.fr}). 


\begin{thebibliography}{73}
\expandafter\ifx\csname natexlab\endcsname\relax\def\natexlab#1{#1}\fi

\bibitem[{{Abel} \& {Satyapal}(2008)}]{AS08}
{Abel}, N.~P., \& {Satyapal}, S. 2008, \apj, 678, 686 

\bibitem[{{Arnaud}(1996)}]{Arnaud96}
{Arnaud}, K.~A. 1996, in Astronomical Society of the Pacific Conference Series,
  Vol. 101, Astronomical Data Analysis Software and Systems V, ed. G.~H.
  {Jacoby} \& J.~{Barnes}, 17

\bibitem[{{Barth} {et~al.}(2005){Barth}, {Greene}, \& {Ho}}]{BGH05}
{Barth}, A.~J., {Greene}, J.~E., \& {Ho}, L.~C. 2005, \apjl, 619, L151

\bibitem[{{Barth} {et~al.}(2004){Barth}, {Ho}, {Rutledge}, \&
  {Sargent}}]{Barth04}
{Barth}, A.~J., {Ho}, L.~C., {Rutledge}, R.~E., \& {Sargent}, W.~L.~W. 2004,
  \apj, 607, 90

\bibitem[{{Barth} {et~al.}(2002){Barth}, {Ho}, \& {Sargent}}]{BHS02}
{Barth}, A.~J., {Ho}, L.~C., \& {Sargent}, W.~L.~W. 2002, \aj, 124, 2607

\bibitem[{{B{\"o}ker} {et~al.}(1997){Barth}, {Forster-Schreiber}, \& {Genzel}}]{Boker97}
{B{\"o}ker}, T., {Forster-Schreiber}, N.~M., \& {Genzel}, R. 1997, \aj, 114, 1883 

\bibitem[{{B{\"o}ker} {et~al.}(2002){B{\"o}ker}, {Laine}, {van der Marel},
  {Sarzi}, {Rix}, {Ho}, \& {Shields}}]{Boker02}
{B{\"o}ker}, T., {Laine}, S., {van der Marel}, R.~P., {Sarzi}, M., {Rix},
  H.-W., {Ho}, L.~C., \& {Shields}, J.~C. 2002, \aj, 123, 1389

\bibitem[{{B{\"o}ker} {et~al.}(2004){B{\"o}ker}, {Sarzi}, {McLaughlin}, {van
  der Marel}, {Rix}, {Ho}, \& {Shields}}]{Boker04}
{B{\"o}ker}, T., {Sarzi}, M., {McLaughlin}, D.~E., {van der Marel}, R.~P.,
  {Rix}, H.-W., {Ho}, L.~C., \& {Shields}, J.~C. 2004, \aj, 127, 105
  
\bibitem[{{B{\"o}ker} {et~al.}(1999){B{\"o}ker}, {van der Marel}, \& {Vacca}}]{Boker99}
{B{\"o}ker}, T., {van der Marel}, R.~P., \& {Vacca}, W.~D. 1999, \aj, 118, 831 

\bibitem[{{Carollo} {et~al.}(1997){Carollo}, {Stiavelli}, {de Zeeuw}, \&
  {Mack}}]{Carollo97}
{Carollo}, C.~M., {Stiavelli}, M., {de Zeeuw}, P.~T., \& {Mack}, J. 1997, \aj,
  114, 2366

\bibitem[{{Carollo} {et~al.}(1998){Carollo}, {Stiavelli}, \&
  {Mack}}]{Carollo98}
{Carollo}, C.~M., {Stiavelli}, M., \& {Mack}, J. 1998, \aj, 116, 68

\bibitem[{{C{\^o}t{\'e}} {et~al.}(2006){C{\^o}t{\'e}}, {Piatek}, {Ferrarese},
  {Jord{\'a}n}, {Merritt}, {Peng}, {Ha{\c s}egan}, {Blakeslee}, {Mei}, {West},
  {Milosavljevi{\'c}}, \& {Tonry}}]{Cote06}
{C{\^o}t{\'e}}, P., {et~al.} 2006, \apjs, 165, 57

\bibitem[{{Dasyra} {et~al.}(2008){Dasyra}, {Ho}, {Armus}, {Ogle}, 
	{Helou}, {Peterson}, {Lutz}, {Netzer}, \& {Sturm}}]{Dasyra08}
{Dasyra}, K.~M., {et~al.} 2008, \apjl, 674, L9 

\bibitem[{{Davidge} \& {Courteau}(2002)}]{DC02}
{Davidge}, T.~J., \& {Courteau}, S. 2002, \aj, 123, 1438 

\bibitem[{{de Vaucouleurs} {et~al.}(1991){de Vaucouleurs}, {de Vaucouleurs},
  {Corwin}, {Buta}, {Paturel}, \& {Fouque}}]{RC3}
{de Vaucouleurs}, G., {de Vaucouleurs}, A., {Corwin}, Jr., H.~G., {Buta},
  R.~J., {Paturel}, G., \& {Fouque}, P. 1991, {Third Reference Catalogue of
  Bright Galaxies} (Berlin Heidelberg New York: Springer-Verlag)

\bibitem[{{Di Matteo} {et~al.}(2000){Di Matteo}, {Quataert}, {Allen},
  {Narayan}, \& {Fabian}}]{DiMatteo00}
{Di Matteo}, T., {Quataert}, E., {Allen}, S.~W., {Narayan}, R., \& {Fabian},
  A.~C. 2000, \mnras, 311, 507
  
\bibitem[{{Dubus} {et~al.}(2004){Dubus}, {Charles}, \& {Long}}]{Dubus04}
{Dubus}, G., {Charles}, P.~A., \& {Long}, K.~S. 2004, \aap, 425, 95 

\bibitem[{{Dubus} \& {Rutledge}(2002)}]{DR02}
{Dubus}, G., \& {Rutledge}, R.~E. 2002, \mnras, 336, 901 

\bibitem[{{Falco} {et~al.}(1999)}]{Falco99}
{Falco}, E.~E., {et~al.} 1999, \pasp, 111, 438 

\bibitem[{{Ferrarese} {et~al.}(2006){Ferrarese}, {C{\^o}t{\'e}}, {Dalla
  Bont{\`a}}, {Peng}, {Merritt}, {Jord{\'a}n}, {Blakeslee}, {Ha{\c s}egan},
  {Mei}, {Piatek}, {Tonry}, \& {West}}]{Ferrarese06}
{Ferrarese}, L., {et~al.} 2006, \apjl, 644, L21

\bibitem[{{Ferrarese} \& {Merritt}(2000)}]{FM00}
{Ferrarese}, L., \& {Merritt}, D. 2000, \apjl, 539, L9

\bibitem[{{Filippenko} \& {Ho}(2003)}]{FH03}
{Filippenko}, A.~V., \& {Ho}, L.~C. 2003, \apjl, 588, L13

\bibitem[{{Filippenko} {et~al.}(1993){Filippenko}, {Ho}, \& {Sargent}}]{FHS93}
{Filippenko}, A.~V., {Ho}, L.~C., \& {Sargent}, W.~L.~W. 1993, \apjl, 410, L75

\bibitem[{{Filippenko} \& {Sargent}(1989)}]{FS89}
{Filippenko}, A.~V., \& {Sargent}, W.~L.~W. 1989, \apjl, 342, L11

\bibitem[{{Fukugita} {et~al.}(1995){Fukugita}, {Shimasaku}, \&
  {Ichikawa}}]{Fukugita95}
{Fukugita}, M., {Shimasaku}, K., \& {Ichikawa}, T. 1995, \pasp, 107, 945

\bibitem[{{Gallo} {et~al.}(2008){Gallo}, {Treu}, {Jacob}, {Woo}, {Marshall}, \&
  {Antonucci}}]{Gallo07}
{Gallo}, E., {Treu}, T., {Jacob}, J., {Woo}, J.-H., {Marshall}, P., \&
  {Antonucci}, R. 2007, \apj, 680, 154

\bibitem[{{Garmire} {et~al.}(2003){Garmire}, {Bautz}, {Ford}, {Nousek}, \&
  {Ricker}}]{Garmire03}
{Garmire}, G.~P., {Bautz}, M.~W., {Ford}, P.~G., {Nousek}, J.~A., \& {Ricker},
  Jr., G.~R. 2003, in Proceedings of the SPIE, Vol. 4851, X-Ray and Gamma-Ray
  Telescopes and Instruments for Astronomy, ed. J.~E. {Truemper} \& H.~D.
  {Tananbaum}, 28

\bibitem[{{Gebhardt} {et~al.}(2000){Gebhardt}, {Bender}, {Bower}, {Dressler},
  {Faber}, {Filippenko}, {Green}, {Grillmair}, {Ho}, {Kormendy}, {Lauer},
  {Magorrian}, {Pinkney}, {Richstone}, \& {Tremaine}}]{Gebhardt00}
{Gebhardt}, K., {et~al.} 2000, \apjl, 539, L13

\bibitem[{{Gebhardt} {et~al.}(2001){Gebhardt}, {Lauer}, {Kormendy}, {Pinkney}, 
{Bower}, {Green}, {Gull}, {Hutchings}, {Kaiser}, {Nelson}, {Richstone}, 
\& {Weistrop}}]{Gebhardt01}
{Gebhardt}, K., {et~al.} 2001, \aj, 122, 2469 

\bibitem[{{Gebhardt} {et~al.}(2002){Gebhardt}, {Rich}, \& {Ho}}]{GRH02}
{Gebhardt}, K., {Rich}, R.~M., \& {Ho}, L.~C. 2002, \apjl, 578, L41

\bibitem[{{Gebhardt} {et~al.}(2005){Gebhardt}, {Rich}, \& {Ho}}]{GRH05}
---. 2005, \apj, 634, 1093

\bibitem[{{Ghez} {et~al.}(2005){Ghez}, {Salim}, {Hornstein}, {Tanner}, {Lu},
  {Morris}, {Becklin}, \& {Duch{\^e}ne}}]{Ghez05}
{Ghez}, A.~M., {Salim}, S., {Hornstein}, S.~D., {Tanner}, A., {Lu}, J.~R.,
  {Morris}, M., {Becklin}, E.~E., \& {Duch{\^e}ne}, G. 2005, \apj, 620, 744

\bibitem[{{Ghosh} {et~al.}(2008){Ghosh}, {Mathur}, {Fiore}, \&
  {Ferrarese}}]{Ghosh08}
{Ghosh}, H., {Mathur}, S., {Fiore}, F., \& {Ferrarese}, L. 2008, \apj, in
  press (arXiv:astro-ph/0801.4382)

\bibitem[{{Gierli{\'n}ski} {et~al.}(2008){Gierli{\'n}ski}, {Niko{\l}ajuk}, \&
  {Czerny}}]{GNC08}
{Gierli{\'n}ski}, M., {Niko{\l}ajuk}, M., \& {Czerny}, B. 2008, \mnras, 383,
  741

\bibitem[{{Greene} \& {Ho}(2004)}]{GH04}
{Greene}, J.~E., \& {Ho}, L.~C. 2004, \apj, 610, 722

\bibitem[{{Greene} \& {Ho}(2005)}]{GH05}
---. 2005, \apj, 630, 122

\bibitem[{{Greene} \& {Ho}(2007{\natexlab{a}})}]{GH07c}
---. 2007{\natexlab{a}}, \apj, 670, 92

\bibitem[{{Greene} \& {Ho}(2007{\natexlab{b}})}]{GH07b}
---. 2007{\natexlab{b}}, \apj, 667, 131

\bibitem[{{Greene} \& {Ho}(2007{\natexlab{c}})}]{GH07a}
---. 2007{\natexlab{c}}, \apj, 656, 84

\bibitem[{{Greene} {et~al.}(2008){Greene}, {Ho}, \& {Barth}}]{GHB08}
{Greene}, J.~E., {Ho}, L.~C., \& {Barth}, A.~J. 2008, \apj, {in press}

\bibitem[{{Ho}(2004)}]{Ho04}
{Ho}, L.~C., ed. 2004, Carnegie Observatories Astrophysics Series, Vol.~1,
  {Coevolution of Black Holes and Galaxies} (Cambridge: Cambridge University
  Press)

\bibitem[{{Ho}(2008)}]{Ho08}
{Ho}, L.~C. 2008, \araa, 46, 475

\bibitem[{{Ho} {et~al.}(2001){Ho}, {Feigelson}, {Townsley}, {Sambruna},
  {Garmire}, {Brandt}, {Filippenko}, {Griffiths}, {Ptak}, \& {Sargent}}]{Ho01}
{Ho}, L.~C., {et~al.} 2001, \apjl, 549, L51

\bibitem[{{Ho} {et~al.}(1995){Ho}, {Filippenko}, \& {Sargent}}]{Ho_2_95}
{Ho}, L.~C., {Filippenko}, A.~V., \& {Sargent}, W.~L. 1995, \apjs, 98, 477

\bibitem[{{Ho} {et~al.}(1997{\natexlab{a}}){Ho}, {Filippenko}, \&
  {Sargent}}]{Ho_3_97}
{Ho}, L.~C., {Filippenko}, A.~V., \& {Sargent}, W.~L.~W. 1997{\natexlab{a}},
  \apjs, 112, 315

\bibitem[{{Ho} {et~al.}(1997{\natexlab{b}}){Ho}, {Filippenko}, \&
  {Sargent}}]{Ho_5_97}
---. 1997{\natexlab{b}}, \apj, 487, 568

\bibitem[{{Iwasawa} {et~al.}(2000){Iwasawa}, {Fabian}, {Almaini}, {Lira},
  {Lawrence}, {Hayashida}, \& {Inoue}}]{Iwasawa00}
{Iwasawa}, K., {Fabian}, A.~C., {Almaini}, O., {Lira}, P., {Lawrence}, A.,
  {Hayashida}, K., \& {Inoue}, H. 2000, \mnras, 318, 879

\bibitem[{{Jarrett} {et~al.}(2000){Jarrett}, {Chester}, {Cutri}, {Schneider},
  {Skrutskie}, \& {Huchra}}]{Jarrett00}
{Jarrett}, T.~H., {Chester}, T., {Cutri}, R., {Schneider}, S., {Skrutskie}, M.,
  \& {Huchra}, J.~P. 2000, \aj, 119, 2498

\bibitem[{{Kennicutt}(1992)}]{Kennicutt92}
{Kennicutt}, Jr., R.~C. 1992, \apj, 388, 310

\bibitem[{{Kormendy}(2004)}]{Kormendy04}
{Kormendy}, J. 2004, in Coevolution of Black Holes and Galaxies, ed. L.~C.
  {Ho}, 1

\bibitem[{{Kormendy} {et~al.}(2008){Kormendy}, {Fisher}, {Cornell}, \&
  {Bender}}]{Kormendy08}
{Kormendy}, J., {Fisher}, D.~B., {Cornell}, M.~E., \& {Bender}, R. 2008, \apjs,
  {submitted}

\bibitem[{{Kormendy} \& {Kennicutt}(2004)}]{KK04}
{Kormendy}, J., \& {Kennicutt}, Jr., R.~C. 2004, \araa, 42, 603

\bibitem[{{Kormendy} \& {McClure}(1993)}]{KM93}
{Kormendy}, J., \& {McClure}, R.~D. 1993, \aj, 105, 1793 

\bibitem[{{La Parola} {et~al.}(2003){La Parola}, {Damiani}, {Fabbiano}, \& {Peres}}]{LaParola03}
{La Parola}, V., {Damiani}, F., {Fabbiano}, G., \& {Peres}, G. 2003, \apj, 583, 758 

\bibitem[{{Leon} \& {Verdes-Montenegro}(2003)}]{LV03}
{Leon}, S., \& {Verdes-Montenegro}, L. 2003, \aap, 411, 391

\bibitem[{{Loewenstein} {et~al.}(2001){Loewenstein}, {Mushotzky}, {Angelini},
  {Arnaud}, \& {Quataert}}]{Loewenstein01}
{Loewenstein}, M., {Mushotzky}, R.~F., {Angelini}, L., {Arnaud}, K.~A., \&
  {Quataert}, E. 2001, \apjl, 555, L21

\bibitem[{{Marconi} \& {Hunt}(2003)}]{MH03}
{Marconi}, A., \& {Hunt}, L.~K. 2003, \apjl, 589, L21

\bibitem[{{McHardy} {et~al.}(2006){McHardy}, {Koerding}, {Knigge}, {Uttley}, \&
  {Fender}}]{McHardy06}
{McHardy}, I.~M., {Koerding}, E., {Knigge}, C., {Uttley}, P., \& {Fender},
  R.~P. 2006, \nat, 444, 730

\bibitem[{{Moran} {et~al.}(2005){Moran}, {Eracleous}, {Leighly}, {Chartas},
  {Filippenko}, {Ho}, \& {Blanco}}]{Moran05}
{Moran}, E.~C., {Eracleous}, M., {Leighly}, K.~M., {Chartas}, G., {Filippenko},
  A.~V., {Ho}, L.~C., \& {Blanco}, P.~R. 2005, \aj, 129, 2108

\bibitem[{Mould} {et~al.}(2000)]{Mould99} 
{Mould}, J.~R., {et~al.} 2000, \apj, 529, {786. Erratum:  2000, \apj, 545, 547}

\bibitem[{{Nagar} {et~al.}(2002){Nagar}, {Falcke}, {Wilson}, \& {Ulvestad}}]{Nagar02}
{Nagar}, N.~M., {Falcke}, H., {Wilson}, A.~S., \& {Ulvestad}, J.~S. 2002, \aap, 392, 53 

\bibitem[{{Nikolajuk} {et~al.}(2004){Nikolajuk}, {Papadakis}, \&
  {Czerny}}]{NPC04}
{Nikolajuk}, M., {Papadakis}, I.~E., \& {Czerny}, B. 2004, \mnras, 350, L26

\bibitem[{{Oliva} {et~al.}(1995){Oliva}, {Origlia}, {Kotilainen}, \&
  {Moorwood}}]{Oliva95}
{Oliva}, E., {Origlia}, L., {Kotilainen}, J.~K., \& {Moorwood}, A.~F.~M. 1995,
  \aap, 301, 55
  
\bibitem[{{Parmar} {et~al.}(2001)}]{Parmar01}
{Parmar}, A.~N., {et~al.} 2001, \aap, 368, 420 
  
\bibitem[{{Paturel} {et~al.}(1999){Paturel}, {Petit}, {Prugniel}, \& {Garnier}}]{Paturel99}
{Paturel}, G., {Petit}, C., {Prugniel}, P., \& Garnier, R. 1999, \aaps, 140, 89 
  
\bibitem[{{Pellegrini}(2005)}]{Pellegrini05}
{Pellegrini}, S. 2005, \apj, 624, 155

\bibitem[{{Peterson}(2007)}]{Peterson07}
{Peterson}, B.~M. 2007, in Astronomical Society of the Pacific Conference
  Series, Vol. 373, The Central Engine of Active Galactic Nuclei, ed. L.~C.
  {Ho} \& J.-W. {Wang}, 3

\bibitem[{{Peterson} {et~al.}(2005){Peterson}, {Bentz}, {Desroches},
  {Filippenko}, {Ho}, {Kaspi}, {Laor}, {Maoz}, {Moran}, {Pogge}, \&
  {Quillen}}]{Peterson05}
{Peterson}, B.~M., {et~al.} 2005, \apj, 632, {799. Erratum: 2006, \apj, 641,
  638}
  
\bibitem[{{Phillips} {et~al.}(1996){Phillips}, {Illingworth}, {MacKenty}, \& {Franx}}]{Phillips96}
{Phillips}, A.~C., {Illingworth}, G.~D., {MacKenty}, J.~W., \& {Franx}, M. 1996, \aj, 111, 1566 

\bibitem[{{Rampazzo} {et~al.}(1995){Rampazzo}, {Reduzzi}, {Sulentic}, \&
  {Madejsky}}]{Rampazzo95}
{Rampazzo}, R., {Reduzzi}, L., {Sulentic}, J.~W., \& {Madejsky}, R. 1995,
  \aaps, 110, 131

\bibitem[{{Remillard} \& {McClintock}(2006)}]{RM06}
{Remillard}, R.~A., \& {McClintock}, J.~E. 2006, \araa, 44, 49

\bibitem[{{Rossa} {et~al.}(2006){Rossa}, {van der Marel}, {B{\"o}ker},
  {Gerssen}, {Ho}, {Rix}, {Shields}, \& {Walcher}}]{Rossa06}
{Rossa}, J., {van der Marel}, R.~P., {B{\"o}ker}, T., {Gerssen}, J., {Ho},
  L.~C., {Rix}, H.-W., {Shields}, J.~C., \& {Walcher}, C.-J. 2006, \aj, 132,
  1074

\bibitem[{{Satyapal} {et~al.}(2007){Satyapal}, {Vega}, {Heckman}, 
{O'Halloran}, \& {Dudik}}]{Satyapal07}
{Satyapal}, S., {Vega}, D., {Heckman}, T., {O'Halloran}, B., 
\& {Dudik}, R. 2007, \apjl, 663, L9 

\bibitem[{{Satyapal} {et~al.}(2008){Satyapal}, {Vega}, {Dudik}, {Abel}, \&
  {Heckman}}]{Satyapal08}
{Satyapal}, S., {Vega}, D., {Dudik}, R.~P., {Abel}, N.~P., \& {Heckman}, T.
  2008, \apj, 677, 926

\bibitem[{{Seth} {et~al.}(2008){Seth}, {Ag{\"u}eros}, {Lee}, \& {Basu-Zych}}]{Seth08} 
{Seth}, A., {Ag{\"u}eros}, M., {Lee}, D., \& {Basu-Zych}, A. 2008, \apj, 678, 116 

\bibitem[{{Seth} {et~al.}(2006){Seth}, {Dalcanton}, {Hodge}, \&
  {Debattista}}]{Seth06}
{Seth}, A.~C., {Dalcanton}, J.~J., {Hodge}, P.~W., \& {Debattista}, V.~P. 2006,
  \aj, 132, 2539

\bibitem[{{Shaw} {et~al.}(1993){Shaw}, {Wilkinson}, \& {Carter}}]{SWC93}
{Shaw}, M., {Wilkinson}, A., \& {Carter}, D. 1993, \aap, 268, 511

\bibitem[{{Sivakoff} {et~al.}(2007){Sivakoff}, {Jord{\'a}n}, {Sarazin},
  {Blakeslee}, {C{\^o}t{\'e}}, {Ferrarese}, {Juett}, {Mei}, \&
  {Peng}}]{Sivakoff07}
{Sivakoff}, G.~R., {et~al.} 2007, \apj, 660, 1246

\bibitem[{{Skrutskie} {et~al.}(2006){Skrutskie}, {Cutri}, {Stiening},
  {Weinberg}, {Schneider}, {Carpenter}, {Beichman}, {Capps}, {Chester},
  {Elias}, {Huchra}, {Liebert}, {Lonsdale}, {Monet}, {Price}, {Seitzer},
  {Jarrett}, {Kirkpatrick}, {Gizis}, {Howard}, {Evans}, {Fowler}, {Fullmer},
  {Hurt}, {Light}, {Kopan}, {Marsh}, {McCallon}, {Tam}, {Van Dyk}, \&
  {Wheelock}}]{2MASS06}
{Skrutskie}, M.~F., {et~al.} 2006, \aj, 131, 1163

\bibitem[{{Soria} {et~al.}(2006){Soria}, {Fabbiano}, {Graham}, {Baldi},
  {Elvis}, {Jerjen}, {Pellegrini}, \& {Siemiginowska}}]{Soria06}
{Soria}, R., {Fabbiano}, G., {Graham}, A.~W., {Baldi}, A., {Elvis}, M.,
  {Jerjen}, H., {Pellegrini}, S., \& {Siemiginowska}, A. 2006, \apj, 640, 126
  
\bibitem[{{Takano} {et~al.}(1994){Takano}, {Mitsuda}, {Fukazawa}, \& {Nagase}}]{Takano94}
{Takano}, M., {Mitsuda}, K., {Fukazawa}, Y., \& {Nagase}, F. 1994, \apjl, 436, L47 

\bibitem[{{Terlevich} {et~al.}(1990){Terlevich}, {Diaz}, \&
  {Terlevich}}]{TDT90}
{Terlevich}, E., {Diaz}, A.~I., \& {Terlevich}, R. 1990, \mnras, 242, 271

\bibitem[{{Tremaine} {et~al.}(2002){Tremaine}, {Gebhardt}, {Bender}, {Bower},
  {Dressler}, {Faber}, {Filippenko}, {Green}, {Grillmair}, {Ho}, {Kormendy},
  {Lauer}, {Magorrian}, {Pinkney}, \& {Richstone}}]{Tremaine02}
{Tremaine}, S., {et~al.} 2002, \apj, 574, 740

\bibitem[{{Tully}(1988)}]{Tully88}
{Tully}, R.~B. 1988, {Nearby Galaxies Catalog} (Cambridge: Cambridge University
  Press)

\bibitem[{{Tully} \& {Shaya}(1984)}]{TS84}
{Tully}, R.~B., \& {Shaya}, E.~J. 1984, \apj, 281, 31

\bibitem[{{Tzanavaris} \& {Georgantopoulos}(2007)}]{TG07}
{Tzanavaris}, P., \& {Georgantopoulos}, I. 2007, \aap, 468, 129

\bibitem[{{V{\'e}ron-Cetty} \& {V{\'e}ron}(2006)}]{VV06}
{V{\'e}ron-Cetty}, M.-P., \& {V{\'e}ron}, P. 2006, \aap, 455, 773

\bibitem[{{Walcher} {et~al.}(2006){Walcher}, {B{\"o}ker}, {Charlot}, {Ho},
  {Rix}, {Rossa}, {Shields}, \& {van der Marel}}]{Walcher06}
{Walcher}, C.~J., {B{\"o}ker}, T., {Charlot}, S., {Ho}, L.~C., {Rix}, H.-W.,
  {Rossa}, J., {Shields}, J.~C., \& {van der Marel}, R.~P. 2006, \apj, 649, 692

\bibitem[{{Walcher} {et~al.}(2005){Walcher}, {van der Marel}, {McLaughlin},
  {Rix}, {B{\"o}ker}, {H{\"a}ring}, {Ho}, {Sarzi}, \& {Shields}}]{Walcher05}
{Walcher}, C.~J., {et~al.} 2005, \apj, 618, 237

\bibitem[{{Wang} {et~al.}(2003){Wang}, {Chaves}, \& {Irwin}}]{WCI03}
{Wang}, Q.~D., {Chaves}, T., \& {Irwin}, J.~A. 2003, \apj, 598, 969 

\bibitem[{{Wehner} \& {Harris}(2006)}]{WH06}
{Wehner}, E.~H., \& {Harris}, W.~E. 2006, \apjl, 644, L17

\bibitem[{{White} \& {Ghosh}(1998)}]{WG98}
{White}, N.~E., \& {Ghosh}, P. 1998, \apjl, 504, L31

\bibitem[{{Whitmore} {et~al.}(1979){Whitmore}, {Schechter}, \&
  {Kirshner}}]{WSK79}
{Whitmore}, B.~C., {Schechter}, P.~L., \& {Kirshner}, R.~P. 1979, \apj, 234, 68

\bibitem[{{Wrobel} {et~al.}(2008){Wrobel}, {Terashima}, \& {Ho}}]{WTH08}
Wrobel, J.~M., Terashima, Y., \& Ho, L. C. 2008, \apj, 675, 1041

\bibitem[{{Yukita} {et~al.}(2007){Yukita}, {Swartz}, {Soria}, \& {Tennant}}]{Yukita07}
{Yukita}, M., {Swartz}, D.~A., {Soria}, R., \& {Tennant}, A.~F. 2007, \apj, 664, 277 

\end{thebibliography}

\clearpage
\pagestyle{empty}
\setlength{\voffset}{25mm}

\begin{deluxetable}{lcccccccccccrrc}
\rotate
\tablewidth{0pt}
\tabletypesize{\scriptsize}
\tablecaption{{\em Chandra} Sample}
\tablehead{\colhead{Galaxy} & \colhead{Distance} & \colhead{$T$} & \colhead{Hubble} & \colhead{Spectral} & \colhead{Nuclear} & \colhead{R.A.} & \colhead{Decl.} & \colhead{Obs. ID} & \colhead{Instrument} & \colhead{Exp. Time} & \colhead{Counts} & \colhead{$\log (L_{\rm X}) $} & \colhead{$\log (L_{\rm lim})$} & \colhead{Sample} \\ 
& \colhead{(Mpc)} & \colhead{Type} & \colhead{Type} & \colhead{Class} & \colhead{Cluster?} & (J2000)& (J2000)& & & \colhead{(ks)}& & \colhead{(ergs s$^{-1}$)} & \colhead{(ergs s$^{-1}$)} & \\
\colhead{(1)} & \colhead{(2)} & \colhead{(3)} & \colhead{(4)} & \colhead{(5)} & \colhead{(6)} & \colhead{(7)} & \colhead{(8)} & \colhead{(9)} & \colhead{(10)} & \colhead{(11)} & \colhead{(12)} & \colhead{(13)} & \colhead{(14)} & \colhead{(15)}}
\startdata
IC\,239 & 14.2 & 6.0 & SAB(rs)cd & L2:: & & 02 36 27.83 & +38 58 08.5 & 7131 & ACIS-S & \phantom{0}4.5 & $<$5 & $<$37.9 & & P \\
IC\,342 & \phantom{0}3.9 & 6.0 & SAB(rs)cd & H & y (B97) & 03 46 48.51& +68 05 45.9 & 6902 & HRC-I & 12.1 & 58 $\pm$ 9 & 37.9 & 37.0 & P \\
IC\,1727 & \phantom{0}6.4 & 9.0 & SB(s)m & T2/L2 & & 01 47 29.88 & +27 20 00.0 & 1634 & ACIS-S & \phantom{0}1.7 & $<$5 & $<$37.7 & & P \\
IC\,3635 & 13.1\tablenotemark{a} & 6.0 & Scd? & & & 12 40 13.33 & +12 52 29.9 & 8104 & ACIS-S & \phantom{0}5.1 & $<$5 & $<$37.8 & & C \\
IC\,5332 & \phantom{0}8.4 & 7.0 & S(s)d & & & 23 34 27.48 &$-$36 06 03.8 & 2067 & ACIS-S & 55.2 & $<$6 & $<$36.4 & & C \\
M\,33 & \phantom{0}0.7 & 6.0 & S(s)cd & H & y (KM93) & 01 33 50.90 & +30 39 35.7 & 3948 & HRC-S & 5.2 & 3638 $\pm$ 61 & 38.9\tablenotemark{b} & 36.3\tablenotemark{b} & P \\
NGC\,45 & \phantom{0}5.9 & 8.0 & S(s)dm & & & 00 14 03.98 &$-$23 10 55.5 & 4690 & ACIS-S & 34.4 & $<$11 & $<$ 36.6 & & C \\
NGC\,55 & \phantom{0}1.3 & 9.0 & SB(s)m: spin & & n (S06) & 00 14 53.60 &$-$39 11 47.8 & 2255 & ACIS-I  & 59.4 & $<$7 & $<$35.1 & & C \\
NGC\,625 & \phantom{0}3.9 & 9.0 & SB(s)m? spin & & & 01 35 04.63 &$-$41 26 10.3 & 4746 & ACIS-S & 60.3 & $<$7 & $<$35.8 & & C \\
NGC\,672 & \phantom{0}7.5 & 6.0 & SB(s)cd & H & & 01 47 54.52 & +27 25 58.0 & 7090 & ACIS-S & \phantom{0}2.2 & $<$5 & $<$37.7 & & P \\
NGC\,925 & \phantom{0}9.4 & 7.0 & SAB(s)d & H & & 02 27 16.91& +33 34 43.9 & 7104 & ACIS-S & \phantom{0}2.2 & 13 $\pm$ 5 & 38.3 & 37.9 & P \\
NGC\,959 & 10.1 & 8.0 & Sdm: & H & & 02 32 23.94 & +35 29 40.8 & 7111 & ACIS-S & \phantom{0}2.2 & $<$5 & $<$37.9 & & P \\
NGC\,988 & 18.3 & 6.0 & SB(s)cd: & & & 02 35 27.72 &$-$09 21 21.6 & 3552 & ACIS-S & \phantom{0}5.3 & $<$5 & $<$38.1 & & C \\
NGC\,1003 & 10.7 & 6.0 & S(s)cd & & & 02 39 16.89 & +40 52 20.2 & 7116 & ACIS-S & \phantom{0}2.7 & $<$5 & $<$37.9 & & C \\
NGC\,1313 & \phantom{0}3.7 & 7.0 & SB(s)d & & n (B02) & 03 18 16.04 &$-$66 29 53.7 & 2950 & ACIS-S & 19.9 & $<$7 & $<$36.2 & & C \\
NGC\,1493 & 11.3 & 6.0 & SB(r)cd & & y (B02) & 03 57 27.38 &$-$46 12 38.6 & 7145 & ACIS-S & 10.0 & 50 $\pm$ 8 & 38.4 & 37.4 & C \\
NGC\,1507 & 10.6 & 9.0 & SB(s)m pec? & & n (P96) & 04 04 27.21\tablenotemark{c} & $-$02 11 18.9\tablenotemark{c} & 7115 & ACIS-S & \phantom{0}2.8 & $<$6 & $<$ 37.9 & & C \\
NGC\,2403 & \phantom{0}4.2 & 6.0 & SAB(s)cd & H & y (DC02) & 07 36 51.39 & +65 36 09.1 & 2014 & ACIS-S & 35.6 & $<$8 & $<$36.1 & & P \\ 
NGC\,2500 & 10.1 & 7.0 & SB(rs)d & H & y (P96) & 08 01 53.22 & +50 44 13.5 & 7112 & ACIS-S & \phantom{0}2.6 &$<$8 & $<$38.1 & & P \\
NGC\,2541 & 10.6 & 6.0 & S(s)cd & T2/H: & & 08 14 40.07 & +49 03 41.1 & 1635 & ACIS-S & \phantom{0}1.9 & $<$5 & $<$38.0 & & P \\
NGC\,2552 & 10.0 & 9.0 & S(s)m? & & y (B02) & 08 19 20.55 & +50 00 35.1 & 7146 & ACIS-S & \phantom{0}7.9 & $<$5 & $<$37.4 & & C \\
NGC\,3079 & 20.4 & 7.0\tablenotemark{d} & SB(s)c spin & S2 & & 10 01 57.92 & +55 40 48.0 & 2038 & ACIS-S & 26.6 & 73 $\pm$ 10 & 38.6 & 37.8 & P \\
NGC\,3184 & \phantom{0}8.7 & 6.0 & SAB(rs)cd & H & & 10 18 16.98 & +41 25 27.7 & 804 & ACIS-S & 42.1 & 133 $\pm$ 13 & 37.9 & 36.6 & P \\
NGC\,3274 & \phantom{0}5.9 & 7.0 & SABd? & & & 10 32 17.28 & +27 40 07.5 & 7085 & ACIS-S & \phantom{0}1.7 & $<$6 & $<$37.7 & & C \\
NGC\,3299 & \phantom{0}5.4 & 8.0 & SAB(s)dm & & & 10 36 23.80 & +12 42 26.6 & 7831 & ACIS-S & \phantom{0}5.1 & $<$5 & $<$37.0 & & C \\
NGC\,3395 & 27.4 & 6.0 & SAB(rs)cd pec: & H & & 10 49 50.11 & +32 58 58.2 & 2042 & ACIS-S & 19.5 & $<$10 & $<$38.2 & & P \\
NGC\,3432 & \phantom{0}7.8 & 9.0 & SB(s)m spin & H & & 10 52 31.13 & +36 37 07.6 & 7091 & ACIS-S & \phantom{0}1.9 & $<$6 & $<$37.8 & & P \\
NGC\,3495 & 12.8 & 7.0 & Sd: & H: & & 11 01 16.16 & +03 37 40.9 & 7126 & ACIS-S & \phantom{0}4.1 & $<$6 & $<$37.9 & & P \\
NGC\,3556 & 14.1 & 6.0 & SB(s)cd spin & H & & 11 11 30.96 & +55 40 26.8 & 2025 & ACIS-S & 59.4 & $<$12 & $<$37.2 & & P \\
NGC\,3985 & \phantom{0}8.3 & 9.0 & SB(s)m: & & & 11 56 42.10 & +48 20 02.2 & 7099 & ACIS-S & \phantom{0}1.7 & $<$6 & $<$37.9 & & C \\
NGC\,4020 & \phantom{0}8.0 & 7.0 & SBd? spin & & & 11 58 56.93 & +30 24 49.0 & 7093 & ACIS-S & \phantom{0}1.7 & $<$5 & $<$37.8 & & C \\
NGC\,4038 & 25.5 & 9.0 & SB(s)m pec & & & 12 01 52.98\tablenotemark{e} &$-$18 52 09.9\tablenotemark{e} & 3043 & ACIS-S & 67.1 & $<$24 & $<$37.9 & & C \\ 
NGC\,4039 & 25.3 & 9.0 & S(s)m pec & & & 12 01 53.64\tablenotemark{e} &$-$18 53 10.8\tablenotemark{e} & 3043 & ACIS-S & 67.1 & 571 $\pm$ 27 & 39.3 & 37.8 & C \\ 
NGC\,4204 & \phantom{0}7.9 & 8.0 & SB(s)dm & & y (B02) & 12 15 14.35 & +20 39 32.6 & 7092 & ACIS-S & \phantom{0}2.0 & $<$6 & $<$37.8 & & C \\
NGC\,4244 & \phantom{0}3.1 & 6.0 & S(s)cd: spin & H & y (S06) & 12 17 29.65 & +37 48 25.5  & 942 & ACIS-S & 49.2 & $<$7 & $<$35.7 & & P \\ 
NGC\,4395 & \phantom{0}4.2\tablenotemark{f} & 9.0 & S(s)m: & S1.8 & y (FH03) & 12 25 48.92 & +33 32 48.2 & 5302 & ACIS-S & 28.2 & 2755 $\pm$ 54 & 39.6\tablenotemark{g} & 37.0\tablenotemark{g} & P \\
NGC\,4411B & 16.8 & 6.0 & SAB(s)cd & & y (B02) & 12 26 47.21 & +08 53 04.4 & 7840 & ACIS-S & \phantom{0}4.9 & $<$5 & $<$38.0 & & C \\
NGC\,4490 & \phantom{0}7.8 & 7.0 & SB(s)d pec & H & & 12 30 36.36 & +41 38 37.0 & 4726 & ACIS-S & 39.6 & 1575 $\pm$ 41 & 38.9 & 36.6 & P \\
NGC\,4559 & \phantom{0}9.7 & 6.0 & SAB(rs)cd & H & & 12 35 57.68 & +27 57 35.1 & 2686 & ACIS-S & \phantom{0}3.0 & 27 $\pm$ 6 & 38.5 & 37.8 & P \\
NGC\,4561 & 12.3 & 8.0 & SB(rs)dm & & & 12 36 08.13 & +19 19 21.3 & 7125 & ACIS-S & \phantom{0}3.5 & 100 $\pm$ 11 & 39.2 & 37.9 & C \\
NGC\,4592 & \phantom{0}9.6 & 8.0 & S(s)dm: & & & 12 39 18.77 & $-$00 31 54.6 & 7105 & ACIS-S & \phantom{0}2.1 & $<$5 & $<$37.9 & & C \\
NGC\,4618 & \phantom{0}7.3 & 9.0 & SB(rs)m & H & y (B02) & 12 41 32.84 & +41 09 02.7 & 7147 & ACIS-S & \phantom{0}9.3 & $<$5 & $<$37.0 & & P \\
NGC\,4625 & \phantom{0}8.2 & 9.0 & SAB(rs)m pec & & y (B02) & 12 41 52.70 & +41 16 25.3 & 7098 & ACIS-S & \phantom{0}1.7 & $<$5 & $<$37.9 & & C \\
NGC\,4631 & \phantom{0}6.9 & 7.0 & SB(s)d spin & H & n (S06) & 12 42 08.00 & +32 32 29.4 & 797 & ACIS-S & 59.2 & $<$8 & $<$36.4 & & P \\
NGC\,4654 & 16.8 & 6.0 & SAB(rs)cd & H & & 12 43 56.63 & +13 07 34.8 & 3325 & ACIS-S & \phantom{0}4.9 & 12 $\pm$ 5 & 38.4 & 38.1 & P \\
NGC\,4688 & 17.1 & 6.0 & SB(s)cd & H & & 12 47 46.46 & +04 20 09.8 & 7859 & ACIS-S & \phantom{0}4.7 & $<$6 & $<$38.1 & & P \\
NGC\,4701 & 20.5 & 6.0 & S(s)cd & & y (B02) & 12 49 11.56 & +03 23 19.4 & 7148 & ACIS-S & \phantom{0}9.0 & 12 $\pm$ 5 & 38.3 & 38.0 & C \\
NGC\,4713 & 17.9 & 7.0 & SAB(rs)d & T2 & & 12 49 57.89 & +05 18 41.1 & 4019 & ACIS-S & \phantom{0}4.9 & 10 $\pm$ 4 & 38.4 & 38.1 & P \\
NGC\,4945 & \phantom{0}5.2 & 6.0 & SB(s)cd: spin & S\tablenotemark{h} & & 13 05 27.27 &$-$49 28 04.4 & 864 & ACIS-S & 49.1 & 379 $\pm$ 22 & 37.9 & 36.7 & C \\
NGC\,5068 & \phantom{0}6.7 & 6.0 & SAB(rs)cd & & y (B02) & 13 18 54.80 &$-$21 02 20.7 & 7149 & ACIS-S & \phantom{0}6.7 & $<$5 & $<$37.1 & & C \\
NGC\,5204 & \phantom{0}4.8 & 9.0 & S(s)m & H & & 13 29 36.50 & +58 25 07.4 & 3933 & ACIS-S & \phantom{0}4.8 & $<$5 & $<$37.0 & & P \\
NGC\,5457 & \phantom{0}5.4 & 6.0 & SAB(rs)cd & H & & 14 03 12.59\tablenotemark{c} & +54 20 56.7\tablenotemark{c} & 2065 & ACIS-S & \phantom{0}9.6 & 20 $\pm$ 6 & 37.3 & 36.8 & P \\ 
NGC\,5474 & \phantom{0}6.0 & 6.0 & S(s)cd pec & H & & 14 05 01.60 & +53 39 43.9 & 7086 & ACIS-S & \phantom{0}1.7 & $<$5 & $<$37.6 & & P \\
NGC\,5585 & \phantom{0}7.0 & 7.0 & SAB(s)d & H & y (B02) & 14 19 48.20 & +56 43 44.5 & 7150 & ACIS-S & \phantom{0}5.3 & $<$5 & $<$37.2 & & P \\
NGC\,5774 & 26.8 & 7.0 & SAB(rs)d & & y (B02) & 14 53 42.75 & +03 34 56.0 & 2940 & ACIS-S & 58.2 & $<$6 & $<$37.5 & & C \\
NGC\,6503 & \phantom{0}6.1 & 6.0 & S(s)cd & T2/S2: & & 17 49 26.51 & +70 08 39.6 & 872 & ACIS-S & 13.2 & 13 $\pm$ 5 & 37.1 & 36.8 & P \\
NGC\,6689 & 12.2 & 7.0 & Sd? spin & H & n (P96) & 18 34 50.24 & +70 31 25.9 & 7123 & ACIS-S & \phantom{0}3.5 & $<$5 & $<$37.9 & & P \\
NGC\,6946 & \phantom{0}5.5 & 6.0 & SAB(rs)cd & H & ? (B02) & 20 34 52.33 & +60 09 13.2 & 1043 & ACIS-S & 58.3 & 766 $\pm$ 29 & 38.2 & 36.2 & P \\
NGC\,7424 & 11.5 & 6.0 & SAB(rs)cd & & y (B02) & 22 57 18.36 & $-$41 04 14.0 & 3496 & ACIS-S & 23.9 & $<$6 & $<$37.1 & & C \\
NGC\,7741 & 12.3 & 6.0 & SB(s)cd & H & n (B02) & 23 43 54.37 & +26 04 32.0 & 7124 & ACIS-S & \phantom{0}3.5 & $<$5 & $<$37.9 & & P \\
NGC\,7793 & \phantom{0}2.8 & 7.0 & S(s)d & & y (B02) & 23 57 49.82 &$-$32 35 27.7 & 3954 & ACIS-S & 48.9 & $<$7 & $<$35.6 & & C \\
\tablebreak
UGC\,05460 & 19.9 & 7.0 & SB(rs)d & & & 10 08 09.30\tablenotemark{i} & +51 50 38.0\tablenotemark{i} & 7835 & ACIS-S & \phantom{0}5.1 & $<$5 & $<$38.2 & & C \\ 
UGC\,08041 & 14.2 & 7.0 & SB(s)d & & & 12 55 12.68 & +00 07 00.0 & 7132 & ACIS-S & \phantom{0}4.7 & $<$5 & $<$37.9 & & C \\
UGC\,09912 & 19.7 & 8.0 & SBdm & & & 15 35 10.50\tablenotemark{j} & +16 32 58.0\tablenotemark{j} & 7832 & ACIS-S & \phantom{0}5.1 & $<$5 & $<$38.2 & & C
\enddata
\tablecomments{Col. (1): Galaxy name. Col. (2): Distance from \cite{Tully88}, assuming a Virgo infall model \citep{TS84} with velocity $v=300$ km s$^{-1}$, $D_{\rm Virgo} = 16.8$ Mpc, and $H_0=75$ km s$^{-1}$ Mpc$^{-1}$, except where noted. Cols. (3)--(4): $T$ type and Hubble type \citep{RC3}. Col. (5): Classification of the nuclear spectrum from \cite{Ho_3_97}, unless otherwise noted: H = \hii \ nucleus, S = Seyfert nucleus, L = LINER, and T = transition object. Quality ratings are given by ``:'' and ``::'' for uncertain and highly uncertain classification, respectively. Two classifications are given for some ambiguous cases, where the first entry corresponds to the adopted choice. Col. (6): Presence of a nuclear star cluster. A ``?'' indicates a non-converging isophotal fit to the surface brightness profile. References are: (B97) \cite{Boker97}; (B02) \cite{Boker02}; (DC02) \cite{DC02}; (FH03) \cite{FH03}; (KM93) \cite{KM93}; (P96) \cite{Phillips96}; (S06) \cite{Seth06}. Cols. (7)--(8): Coordinates from 2MASS \citep{2MASS06}, unless otherwise noted. Col. (9): {\em Chandra} observation ID. Col. (10): Instrument used. Col. (11): Live exposure time of observation. Col. (12): Background-subtracted counts measured, or upper limit if no source was detected. Col. (13): 2--10 keV X-ray luminosity or upper limit if no source was detected, assuming a power-law spectrum with index $\Gamma = 1.8$ and Galactic column density of $N_{\rm H} = 2 \times 10^{20}$ cm$^{-2}$ \citep{Ho01}, unless otherwise noted. Col. (14): Limiting 2--10 keV luminosity for observations with a detected source. Col. (15): ``P'' object is in {\em Chandra} archive and within the Palomar survey of \cite{Ho_2_95}; ``C'' object is within {\em Chandra} archive but not part of the Palomar survey.}
\tablenotetext{a}{ Distance from \cite{Mould99}, using the Virgo infall only model.}
\tablenotetext{b}{ Assuming a column density of $N_{\rm H} = 2 \times 10^{21}$ cm$^{-2}$ \citep{Takano94,Parmar01,DR02,LaParola03}.}
\tablenotetext{c}{ Optical position from \cite{LV03}, with a $1\sigma$ uncertainty of 0\farcs5.}
\tablenotetext{d}{ NGC\,3079 is listed in \cite{RC3} as having a $T$ type of 7.0 and a morphological type of SB(s)c spin, which is inconsistent. For this study, we assume that it is of a type later than SBc, and include it our sample.}
\tablenotetext{e}{ Optical nuclear position determined from NED, with a $1\sigma$ uncertainty of 1\arcsec.}
\tablenotetext{f}{ Distance from \cite{FH03}.}
\tablenotetext{g}{ Assuming a column density of $N_{\rm H} = 2 \times 10^{22}$ cm$^{-2}$ \citep{Iwasawa00,Moran05}.}
\tablenotetext{h}{ Classification from \cite{VV06}.}
\tablenotetext{i}{ Optical nuclear position from \cite{Falco99}, with a $1\sigma$ uncertainty of 1\arcsec.}
\tablenotetext{j}{ Optical nuclear position from \cite{Paturel99}, with a $1\sigma$ uncertainty of 1\arcsec.}

\label{tab:log}
\end{deluxetable}
\clearpage

\clearpage
\pagestyle{plaintop}
\setlength{\voffset}{0mm}
\begin{deluxetable}{lllcccccrl}
\rotate
\tablewidth{0pt}
\tabletypesize{\scriptsize}
\tablecaption{AGN Candidates}
\tablehead{ \colhead{Galaxy} & \colhead{$\delta$} & \colhead{$\delta$} & \colhead{$\sigma_\star$} & \colhead{$\log M_{\rm BH}$} & \colhead{$\log L_{\rm X} $} & \colhead{$\log L_{\rm X} / L_{\rm Edd} $} & \colhead{$H_1$} & \colhead{$H_2$} & \colhead{Notes} \\
 & \colhead{(\arcsec)} & \colhead{(pc)} & \colhead{(km s$^{-1}$)} & \colhead{($M_\odot$)} & \colhead{(ergs s$^{-1}$)} & & & & \\
\colhead{(1)} & \colhead{(2)} & \colhead{(3)} & \colhead{(4)} & \colhead{(5)} & \colhead{(6)} & \colhead{(7)} & \colhead{(8)} & \colhead{(9)} & \colhead{(10)}} 
\startdata
IC\,342 & 1.96 & \phantom{0}37.0 & 33 $\pm$ 3\tablenotemark{a} (1) & 6.5 & 37.9 &$-$6.7 &  & & colors N/A \\
NGC\,925 & 1.65 & \phantom{0}75.3 & & & 38.3 & & &  & too few counts for colors \\
NGC\,1493 & 0.68 & \phantom{0}37.4 & 25.0 $\pm$ 3.8\tablenotemark{a} (2) & 4.5 & 38.4 &$-$4.3 & 0.08 $\pm$ 0.16 &$-$0.31 $\pm$ 0.15 & \\
NGC\,3079 & 1.29 & 127.6 & 150 $\pm$ 10 (3) & 7.6 & 38.6 &$-$7.1 &  & 0.35 $\pm$ 0.15 & bright diffuse emission; \\
& & & & & & & & & no soft source; known Seyfert 2 \\
NGC\,3184 & 2.19 & \phantom{0}92.1 & & & 37.9 & & 0.09 $\pm$ 0.10 &$-$0.34 $\pm$ 0.09 & double source? \\
NGC\,4039 & 2.08\tablenotemark{b} & 255.0 & & & 39.3 & & 0.01 $\pm$ 0.04 &$-$0.30 $\pm$ 0.04 & \\
NGC\,4395 & 1.08 & \phantom{0}22.0 & $<$30\tablenotemark{a} (4) & $<$4.8 & 39.6 & $< -$3.3 & 0.06 $\pm$ 0.01 & 0.74 $\pm$ 0.02 & known Seyfert 1\\
NGC\,4490 & 1.91 & \phantom{0}72.2 & & & 38.9 & & 0.47 $\pm$ 0.02 &$-$0.16 $\pm$ 0.03 \\
NGC\,4559 & 1.03 & \phantom{0}48.7 & & & 38.5 & & &$-$0.42 $\pm$ 0.29 & no soft source\\
NGC\,4561 & 1.60 & \phantom{0}95.5 & & & 39.2 & & 0.07 $\pm$ 0.10 &$-$0.16 $\pm$ 0.10 \\
NGC\,4654 & 1.59 & 129.8 & & & 38.4 & & & & too few counts for colors \\
NGC\,4701 & 2.92 & 290.5 & & & 38.3 & & & & too few counts for colors \\
NGC\,4713 & 0.30 & \phantom{0}26.3 & & & 38.4 & & & & too few counts for colors\\
NGC\,4945 & 3.17 & \phantom{0}79.8 & 134 $\pm$ 20 (5) & 7.4 & 37.9 &$-$7.7 & & & bright diffuse emission; no soft \\
& & & & & & & & & or medium source; known Seyfert \\
NGC\,5457 & 0.98\tablenotemark{c} & \phantom{0}25.6 & 80 $\pm$ 20 (6) & 6.5 & 37.3 &$-$7.3 & & & too few counts for colors\\
NGC\,6503 & 0.95 & \phantom{0}28.1 & 46 $\pm$ 3 (8) & 5.6 & 37.1 &$-$6.5 & & & too few counts for colors; \\
& & & & & & & & & known nuclear activity \\
NGC\,6946 & 1.28 & \phantom{0}34.2 & & & 38.2 & & 0.47 $\pm$ 0.04 &$-$0.22 $\pm$ 0.04 & double source? \\
\vspace{-6pt}
\enddata
\tablecomments{Col. (1): Name of AGN candidate. Col. (2): Offset, in arcseconds, between the 2MASS nuclear position and the centroid of the X-ray source, except where noted. Col. (3): Same as Col. (2), except in pc. Col. (4): Stellar velocity dispersion of galaxy, unless otherwise noted. References are: (1) \cite{Boker99}; (2) \cite{Walcher05}; (3) \cite{SWC93}; (4) \cite{FH03}; (5) \cite{Oliva95}; (6) \cite{WSK79}; (7) \cite{Rampazzo95}; (8) \cite{BHS02}. Col. (5): Estimate of the BH mass using $\log (M_{\rm BH}/M_\odot) = 8.13 + 4.02 \log (\sigma / 200 {\rm \ km \ s}^{-1})$ \citep{Tremaine02}. Col. (6): X-ray luminosity reproduced from Table~\ref{tab:log}. Col (7): Ratio of X-ray luminosity to Eddington luminosity. Col. (8): Soft X-ray color, as defined in the text. Col. (9): Hard X-ray color, as defined in the text. Col. (10): Notes on individual sources.}
\tablenotetext{a}{ Velocity dispersion of nuclear star cluster.}
\tablenotetext{b}{ Optical nuclear position determined from NED, with a $1\sigma$ uncertainty of 1\arcsec.}
\tablenotetext{c}{ Optical position from \cite{LV03}, with a $1\sigma$ uncertainty of 0\farcs5.}
\label{tab:agn}
\end{deluxetable}
\clearpage

\clearpage

\clearpage
\pagestyle{plaintop}
\setlength{\voffset}{0mm}
\begin{deluxetable}{lccccccc}
\tablewidth{0pt}
\tabletypesize{\scriptsize}
\tablecaption{Spectral Fits}
\tablehead{\colhead{Galaxy} & \colhead{$N_{\rm H}$} & \colhead{$\Gamma$} & \colhead{Norm$_{\rm pl}$} & \colhead{$kT$} & \colhead{Norm$_{\rm bb}$} & \colhead{$\chi^2$} & \colhead{dof} \\
& \colhead{($\times 10^{22}$ cm$^{-2}$)} & & \colhead{(cts s$^{-1}$ keV$^{-1}$)} & \colhead{(keV)} &\colhead{(cts s$^{-1}$ keV$^{-1}$)} \\
\colhead{(1)} & \colhead{(2)} & \colhead{(3)} & \colhead{(4)} & \colhead{(5)} & \colhead{(6)} & \colhead{(7)} & \colhead{(8)} \\
}
\startdata
NGC\,4039 \\
\ \ \ abs. power law & 0.20 $\pm$ 0.05 & 2.7 $\pm$ 0.2 & $(1.8 \pm 0.6) \times 10^{-5}$ & \ldots & \ldots & 1.38 & 24 \\
\ \ \ bbody & \ldots & \ldots & \ldots & 0.35 $\pm$ 0.02 & $(3.4 \pm 0.3) \times 10^{-7}$ & 2.54 & 25 \\
\ \ \ abs. pl + bbody & 0.7 $\pm$ 0.2 & 2.8 $\pm$ 0.4 & $(3 \pm 2) \times 10^{-5}$ & 0.09 $\pm$ 0.01& $(0.5 \pm 2) \times 10^{-4}$ & 0.83 & 22
\vspace{0.1cm} \\
\hline 
\vspace{-0.2cm} \\
NGC\,4395 \\
\ \ \ abs. power law & 2.7 $\pm$ 0.2 & 0.8 $\pm$ 0.1 & $(2.0 \pm 0.6) \times 10^{-4}$ & \ldots & \ldots & 1.15 & 110 \\
\ \ \ bbody & \ldots & \ldots & \ldots & 4.5 $\pm$ 0.3 & $(1.8 \pm 0.6) \times 10^{-4}$ & 3.36 & 111
\vspace{0.1cm} \\
\hline 
\vspace{-0.2cm} \\
NGC\,4490 \\
\ \ \ abs. power law & 0.80 $\pm$ 0.06 & 2.3 $\pm$ 0.1 & $(1.9 \pm 0.5) \times 10^{-4}$ & \ldots & \ldots & 1.10 & 60 \\
\ \ \ bbody & \ldots & \ldots & \ldots & 0.78 $\pm$ 0.02 & $(3.4 \pm 0.3) \times 10^{-6}$ & 1.56 & 61 
\vspace{0.1cm} \\
\hline 
\vspace{-0.2cm} \\
NGC\,6946 \\
\ \ \ abs. power law & 0.50 $\pm$ 0.09 & 1.9 $\pm$ 0.2 & $(4 \pm 1) \times 10^{-5}$ & \ldots & \ldots & 0.59 & 30 \\
\ \ \ bbody & \ldots & \ldots & \ldots & 0.73 $\pm$ 0.03 & $(1.1 \pm 0.1) \times 10^{-6}$ & 0.73 & 31
\enddata
\tablecomments{Col. (1): Name of AGN candidate and model used in fit. Col. (2): Neutral column density. Col. (3): Power-law index (where $N(E) \propto E^{-\Gamma}$). Col. (4): Normalization of power-law component. Col. (5): Temperature of blackbody component. Col. (6): Normalization of blackbody component. Col (7): Reduced $\chi^2$. Col. (8): Degrees of freedom.}
\label{tab:spectra}
\end{deluxetable}
\clearpage

\clearpage

\clearpage
\epsscale{0.9}
{\plotone{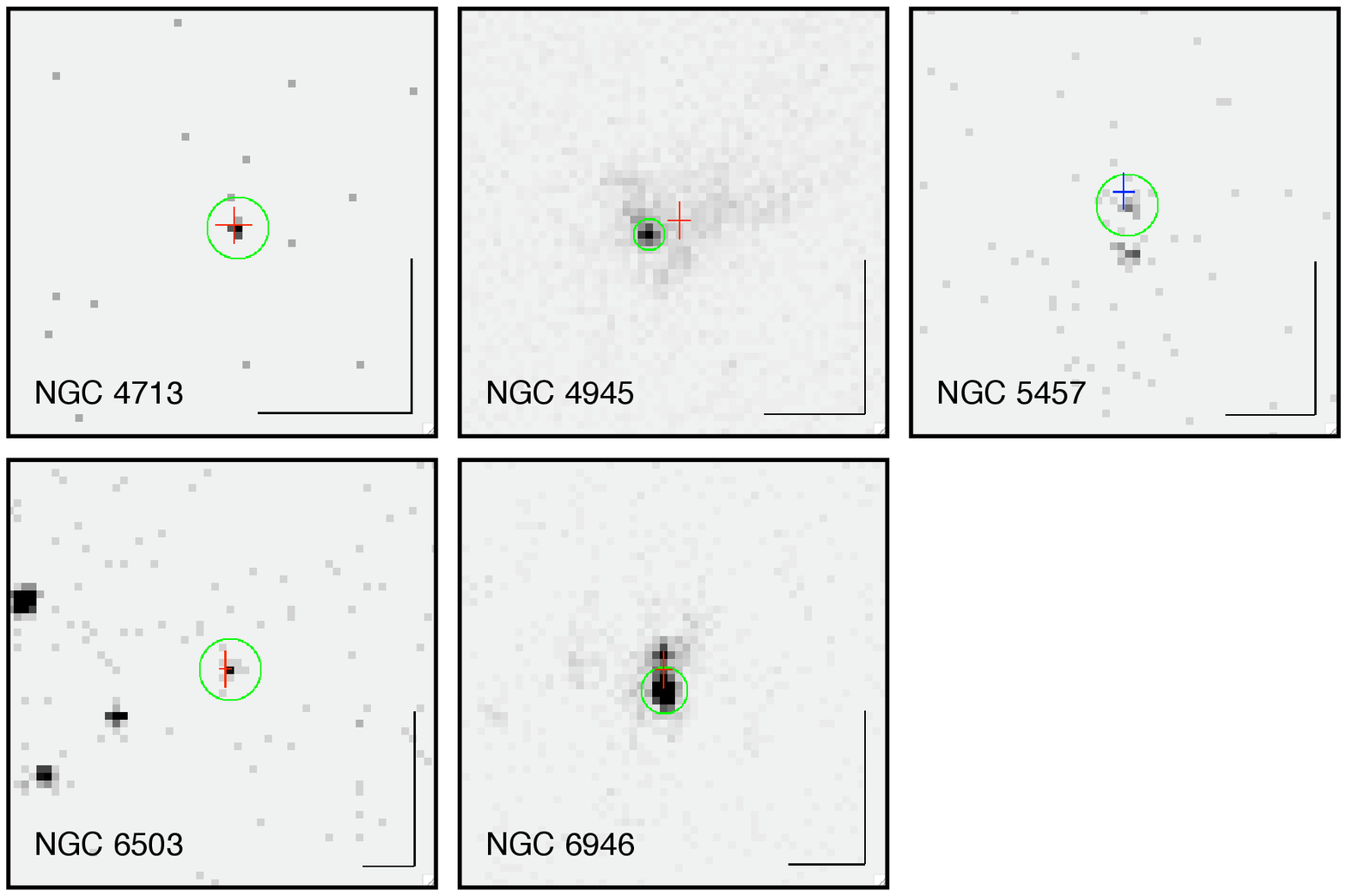}}\\
\centerline{Fig. 2. --- Continued.}

\clearpage

\clearpage

\clearpage

\end{document}